\documentclass[twocolumn]{aastex62}

\usepackage{amsmath, natbib}

\shorttitle{Connection between AGN variability and black hole physical properties}
\shortauthors{S\'anchez-S\'aez et al.}

\begin{document}

\title{The QUEST-La Silla AGN variability survey: connection between AGN variability and black hole physical properties}

\author{P. S\'anchez-S\'aez}
\affiliation{Departamento de Astronomía, Universidad de Chile, Casilla 36D, Santiago, Chile}
\affiliation{European Southern Observatory, Casilla 19001, Santiago 19, Chile}

\author{P. Lira}
\affiliation{Departamento de Astronomía, Universidad de Chile, Casilla 36D, Santiago, Chile}

\author{J. Mej\'ia-Restrepo}
\affiliation{European Southern Observatory, Casilla 19001, Santiago 19, Chile}
\affiliation{Departamento de Astronomía, Universidad de Chile, Casilla 36D, Santiago, Chile}

\author{L. C. Ho}
\affiliation{Kavli Institute for Astronomy and  Astrophysics, PekingUniversity, Beijing 100871, China}
\affiliation{Department of Astronomy, School of Physics, Peking University, Beijing 100871, China}

\author{P. Ar\'evalo}
\affiliation{Instituto de Física y Astronomía, Facultad de Ciencias, Universidad de Valparaíso, Gran Bretana No. 1111, Playa Ancha, Valparaíso, Chile}

\author{M. Kim}
\affiliation{Korea Astronomy and Space Science Institute, Daejeon 305-348, Republic of Korea}
\affiliation{Department of Astronomy and Atmospheric Sciences, Kyungpook National University, Daegu  702-701, Korea}

\author{R. Cartier}
\affiliation{Cerro Tololo Inter-American Observatory, National Optical Astronomy Observatory, Casilla 603, La Serena, Chile}

\author{P. Coppi}
\affiliation{Yale Center for Astronomy and Astrophysics, 260 Whitney Avenue, New Haven, CT 06520, USA}

%% Mark off the abstract in the ``abstract'' environment. 
\begin{abstract}

We present our statistical analysis of the connection between active galactic nuclei (AGN) variability and physical properties of the central supermassive black hole (SMBH). We constructed optical light curves using data from the QUEST-La Silla AGN variability survey. To model the variability, we used the structure function, among the excess variance and the amplitude from Damp Random Walk (DRW) modeling. For the measurement of SMBH physical properties, we used public spectra from the Sloan Digital Sky Survey (SDSS). Our analysis is based on an original sample of 2345 sources detected in both SDSS and QUEST-La Silla. For 1473 of these sources we could perform a proper measurement of the spectral and variability properties, and 1348 of these sources were classified as variable ($91.5\%$). We found that the amplitude of the variability ($A$) depends solely on the rest frame emission wavelength and the Eddington ratio, where $A$ anti-correlates with both $\lambda_{rest}$ and $L/L_{\text{Edd}}$. This suggests that AGN variability does not evolve over cosmic time, and its amplitude is inversely related to the accretion rate. We found that the logarithmic gradient of the variability ($\gamma$) does not correlate significantly with any SMBH physical parameter, since there is no statistically significant linear regression model with an absolute value of the slope higher than 0.1. Finally, we found that the general distribution of $\gamma$ measured for our sample differs from the distribution of $\gamma$ obtained for light curves simulated from a DRW process. For 20.6\% of the variable sources in our sample, a DRW model is not appropriate to describe the variability, since $\gamma$ differs considerably from the expected value of 0.5.

\end{abstract}


\section{Introduction}\label{intro}

Active Galactic Nuclei (AGN) show time-variable emission in every waveband in which they have been studied. The characteristic time-scales of the variability range from hours to years, with the shortest time-scales being associated with shorter emission wavelengths. This can be understood in the context of the current AGN structure models, where ultraviolet (UV) and optical emission are originated in an accretion disk around a super-massive black hole (SMBH), and non-thermal X-ray emission is produced in a inner hot plasma component (corona), which is geometrically much smaller and more concentrated than the accretion disk, and therefore able to show more rapid variability. Intensive monitoring of nearby AGN suggests that short term variability from the UV to the near-IR could be driven by the rapid changes in the X-ray flux, which illuminates the accretion disk producing the short term UV/optical variations, since small lags between optical and X-ray bands have been found in reverberation mapping (RM) analyses. However, it has been noticed that at time-scales of months or years, the amplitude of the UV/optical variability is larger than the amplitude of the X-ray variability, which implies that X-ray reprocessing is not the main source of the UV/optical variations, and intrinsic variability from the accretion disk is required \citep{Krolik91,Arevalo08,Lira15,Edelson15}.

Even though variability is one of the defining characteristics of AGN we do not completely
understand the mechanisms that drive such variations. In particular it is not clear yet how physical properties of the central engine (e.g., luminosity, black hole mass, Eddington ratio, etc) are related to variability properties of the system (e.g., characteristic time-scale, variability amplitude, etc). If we can establish a firm statistical correlation between certain AGN variability features and some SMBH physical properties, we will be able to use the variability as a tool in the future to derive physical properties for huge samples of objects from dedicated synoptic surveys such as the Large Synoptic Survey Telescope (LSST; \citealt{LSST}). Several efforts have been made in the past to assess this issue, some of them restricting the analysis to small numbers of well sampled light curves (e.g. \citealt{Kelly09,Kelly13,Simm16,Smith18}), or studying large samples of sources through ensemble light curve analysis, assuming that sources with similar physical properties would have similar variability features (e.g. \citealt{Wilhite08,Bauer09,MacLeod10,Caplar17}). In order to test whether this assumption is correct, we need to perform an analysis of well sampled individual AGN light curves, with known physical properties. Hence, long and intensive campaigns are crucial.

An anti-correlation between the amplitude of the UV-optical variability and luminosity has been consistently observed by previous studies (e.g \citealt{Angione72,Hook94,Cristiani97,VandenBerk04,Wilhite08,Bauer09,Kelly09,MacLeod10,Kelly13,Simm16,Caplar17}). However, the existence of correlation between the amplitude of the variability and the black hole mass or the Eddington ratio is not clear yet. \cite{Wold07} used a sample of $\sim 100$ quasars from the Quasar Equatorial Survey Team, Phase 1 (QUEST1) variability survey \citep{Rengstorf04}. They found a positive correlation between the black hole mass and the amplitude of the variability. \cite{Wilhite08} found a positive correlation between the amplitude of the variability with black hole mass, and proposed that this could be explained by an anti-correlation with the Eddington ratio. \cite{MacLeod10} also found a positive correlation with black hole mass, and propose that the anti-correlation between the amplitude of the variability and the Eddington ratio exists, but an additional dependence on luminosity or black hole mass is required. \cite{Kelly09} found no evidence of correlation between the amplitude of the variability and the black hole mass or the Eddington ratio, and \cite{Kelly13} found a scattered correlation between the amplitude and the black hole mass, and a weak anti-correlation with the Eddington ratio. \cite{Simm16} found no correlation with the black hole mass, and an anti-correlation with Eddington ratio. More recently, \cite{Li18} used a large sample of quasars ($\sim 10^5$) to perform an ensemble variability analysis. They found that the amplitude of the variability correlates positively with redshift, and negatively with bolometric luminosity, rest-frame wavelength and Eddington ratio. They also found that the correlation with black hole mass was uncertain. This uncertainty can be produced by the use of ensemble light curves and also by the large uncertainties that might be present in the black hole mass estimations used in their analysis (taken from \citealt{Kozlowski17a}), since they are calculated by using luminosities derived from broadband extinction-corrected magnitudes obtained from the Sloan Digital Sky Survey (SDSS; \citealt{York00}), and by using the full width at half maximum (FWHM) of the lines obtained by \cite{Paris17a}. It is clear that all these results on the correlation with black hole mass and Eddington ratio are inconsistent, most likely due to the shortcomings on the samples used, as highlighted before.

\cite{Rakshit17} used a large sample of narrow-line Seyfert 1 (NLSy1) and broad-line Seyfert 1 (BLSy1) from the Catalina Real Time Transient Survey (CRTS; \citealt{Drake09}). The light curves used in their analysis have a minimum of 50 epochs of data spanning 5 to nine years, thus they could perform a variability analysis for individual light curves. They found a strong anti-correlation between the amplitude of variability and the Eddington ratio, and they proposed that the accretion disk is the main driver of the variability observed in both broad and narrow line Seyfert 1 galaxies. However, since \cite{Rakshit17} used Damp Random Walk (DRW) modelling to measure the variability amplitude, which has several limitations for the analysis of ground-based light curves, since they tend to have gaps and time coverages of a few months or years (see section \ref{var_features}), their results must be confirmed using a different method (e.g. the structure function).

Between 2010 and 2015 we carried out an AGN variability survey using
the wide-field QUEST camera on the 1m ESO-Schmidt telescope at La
Silla Observatory, observing five extragalactic fields: Stripe82,
Elais-S1, COSMOS, ECDFS and XMM-LSS. 
These are some the most
intensively observed regions in the sky, with a huge amount of
ancillary data ranging from X-rays to radio waves. The aims of our survey are: 1) to test and improve variability
selection methods of AGN, and find AGN populations missed by other
optical selection techniques \citep{Schmidt10,Butler11,PalanqueDelabrouille11}, which is the subject of a forthcoming paper; 2) to obtain a large number of well--sampled light curves, covering time-scales ranging from days
to years; 3) to study the link between the variability properties
(e.g., characteristic time-scales and amplitudes of variation) with
physical parameters of the system (e.g., black-hole mass, luminosity,
and Eddington ratio). \cite{Cartier15} presented
the technical description of the survey, the full characterisation of
the QUEST camera, and a study of the relation of variability with
multi-wavelength properties of X-ray selected AGN in the COSMOS
field.

In this paper we present our statistical analysis of the connection between AGN variability and physical properties of SMBH. For the variability analysis we used light curves from the QUEST-La Silla AGN variability survey, and derived physical properties from spectra taken from SDSS. We perform the spectral fitting using the procedure of \cite{MejiaRestrepo16} (MR16 hereafter), from which we could derive physical parameters and also line fitting properties such as the FWHM of the emission lines and continuum luminosities. For the variability analysis, we used the same approach as in \cite{Sanchez17} (S17 hereafter). In this work, we used single object light curves, in order to test the claim that sources with similar physical properties have similar variability behaviors (like proposed by \citealt{VandenBerk04,Wilhite08,MacLeod10,Caplar17}, among others). 

The paper is organized as follows. In section \ref{data} we describe the optical imaging and spectroscopic data used for the analysis. In section \ref{var_analysis} we describe the different variability features used, and we report the results of the variability analysis for our sample. In section \ref{spec_analysis} we explain the procedure followed to obtain the physical properties from the SDSS spectra, and show the distribution of these parameters for our sample. In section \ref{sample} we define the different sub-samples used in our analysis. In section \ref{var_vs_spec} we show the results of our statistical analysis done to connect the variability and physical properties. In section \ref{var_classes} we analyse the differences in the variability parameters of sources classified as Broad Line QSO and normal sources, and sources classified as radio-loud and radio-quiet. Finally, in section \ref{discussion} we discuss the physical implications of our findings and summarize the main results. The photometry reported here is in the AB system. We adopt the cosmological parameters $H_0=70$ km s$^{-1}$ Mpc$^{-1}$, $\Omega_m=0.3$ and $\Omega_\Lambda=0.7$.

\section{Data}\label{data}

\subsection{Optical light curves}\label{light_curves}

We reduced the data from the QUEST--La Silla AGN variability survey (hereafter QUEST) using our own customized pipeline, following the same procedure described by \cite{Cartier15}. The survey uses a broadband filter, the Q-band, similar to the union of the g and r SDSS filters. Our QUEST fields are much bigger than just COSMOS, ELAIS, etc., even though we use the same names for them, with a surveyed area of $\sim 7$ deg$^2$ per field. One of the advantages of our survey over other surveys was the very intense monitoring, observing the fields every possible night (although with large observing gaps from 2010 to 2012 due to telescope failures). In average we obtained between 2 to 5 observations per night, per every observable field. Individual images reached a limiting magnitude between $r \sim 20.5$ and $r \sim 21.5$ mag for a exposure time of 60 seconds or 180 seconds, respectively.

To calibrate the photometry, we used public photometric SDSS catalogs \citep{Gunn98,Doi10} for the COSMOS, Stripe82 and XMM-LSS fields, and public catalogs from the Dark Energy Survey (DES; \citealt{Abbott18}) for the ELAIS-S1 and ECDFS fields. We then constructed light curves for all the sources from the SDSS and DES catalogs with detections in the QUEST data, using the same methodology as in \cite{Cartier15}. We decided to bin our light curves every three days, in order to reduce the noise in our light curves, produced by changes in atmospheric conditions, the relatively low quality of the QUEST camera, among other factors. We generated a total of $\sim 450.000$ binned light curves.

\subsection{SDSS spectra}\label{spectra}

Three of our fields (COSMOS, Stripe82 and XMM-LSS) have spectroscopic information from the SDSS survey. We used the SDSS Data Release 14 Quasar catalog (DR14Q) \citep{Paris17b}, in order to identify sources with a detection in QUEST already classified as quasars. We found 2345 sources with both QUEST light curves and SDSS spectra, classified as quasars in DR14Q,  this sample inherits the selection criteria of SDSS spectroscopic survey. We downloaded the calibrated SDSS spectra from the SDSS Catalog Archive Server, and then corrected the spectra by Galactic extinction using the maps of \cite{Schlegel98} and the model of \cite{Cardelli89}. The wavelength coverage of the SDSS spectra ranges from 3800 to 9200 {\AA} for the SDSS survey and from 3650 to 10400 {\AA} for the BOSS survey \citep{Dawson13}, with a spectral resolution of 1500 at 3800 \AA, and 2500 at 9000 \AA.

From the SDSS quasar catalogs we obtained the spectroscopic redshift ($z$) for every source. We then used these redshifts to transform every light curve to the AGN rest frame: $t_{\text{rest}} = t_{\text{obs}}/(1 + z)$, where $t_{\text{rest}}$ is the light curve time at the rest frame in days and $t_{\text{obs}}$ is the observed time. The following analysis has been done considering the light curves in rest frame time.

\section{Variability analysis}\label{var_analysis}

\subsection{Variability features}\label{var_features}

To characterize the variability of our sources, we used the same approach as S17. We used two parameters related to the amplitude of the variability: $P_{var}$ and the excess variance ($\sigma_{rms}$), and two methods related to the structure of the variability: the Structure Function and the Damped Random Walk process (DRW). Here we describe briefly every feature. For further details see S17 and references therein.

$P_{var}$ \citep{McLaughlin96,Paolillo04,Young12,Lanzuisi14,Cartier15,Sanchez17} corresponds to the probability that the source is intrinsically variable. It considers the $\chi^2$ of the light curve, and calculates the probability $P_{var}=P(\chi^2)$ that a $\chi^2$ lower or equal to the observed value could occur by chance for an intrinsically non-variable source.

$\sigma_{rms}$ \citep{Nandra97,Turner99,Allevato13,Lanzuisi14,Cartier15,Simm16,Sanchez17} is a measure of the intrinsic variability amplitude. It is calculated as $\sigma^2_{rms}=(\sigma_{LC}^2-\overline{\sigma}_{m}^2)/\overline{m}^2$, where $(\sigma_{LC}$ is the standard deviation of the light curve, $\overline{\sigma}_{m}$ is the mean photometric error, and $\overline{m}$ is the mean magnitude. We can associate an error for $\sigma_{rms}$ due to Poisson noise, $err(\sigma^2_{rms})=S_D/(\bar{x}^2 N_{obs}^{1/2})$, where $$S^2_D=\frac{1}{N_{obs}}\sum^{N_{obs}}_{i=1}\{[(x_i-\bar{x})^2-\sigma^2_{err,i}]-\sigma^2_{rms}\bar{x}^2\}^2.$$

Following S17, we classify a source as variable if its light curve satisfies $P_{var} \geq 0.95$ and $(\sigma^2_{rms}-err(\sigma^2_{rms}))>0$.

\citet{Kelly09} proposed that a DRW process can be a good descriptor for AGNs light curves. This process model a light curve with a stochastic differential equation that includes a damping term that pushes the signal back to its mean: $dX(t)=-\frac{1}{\tau}X(t)dt+\sigma_{DRW}\sqrt{dt}\,\epsilon(t)+b\,dt$  with $\tau,\sigma_{DRW},t>0$. $\tau$ corresponds to the ``relaxation time" of the process or the characteristic time for the time series to become roughly uncorrelated, and has units of days, $\sigma_{DRW}$ corresponds to the amplitude of the variability at short time-scales ($t<<\tau$), and has units of mag/day$^{1/2}$. The long time-scale variability (SF$_{\text{inf}}$) is calculated as $\sigma_{DRW}\sqrt{\tau/2}$. \cite{Kozlowski17} and S17 demonstrated the limitations of the use of DRW processes for short light curves. For light curves with $t_{rest}<10\times\tau$, the correct value of $\tau$ cannot be determined. Since our light curves have an observed time coverage of $t_{obs}\leqslant 5$ years, while the characteristic time-scale is expected to be of the order of hundred of days, we decided to excluded $\tau$ from our analysis.

The Structure Function (SF) \citep{Cristiani96,Giveon99,VandenBerk04, deVries05,Rengstorf06,Schmidt10,PalanqueDelabrouille11,Graham14,Cartier15,Kozlowski16a,Caplar17,Sanchez17} is a measure of the amplitude of the variability as a function of the time lapse between compared observations ($\tau$). There are several definitions in the literature for SF (see \citealt{Kozlowski16a} for a good summary), however S17 demonstrated that the best definition of SF for irregularly sampled and noisy light curves is the bayessian definition of \cite{Schmidt10}. They model the structure function with a power-law using a Markov Chain Monte Carlo (MCMC) method, where $\text{SF}(\tau)=A\left( \frac{\tau}{1\text{yr}}\right)^{\gamma} $. In this case $A$ corresponds to the amplitude of the variability at 1 year in the rest frame, and $\gamma$  is the logarithmic gradient of this change in magnitude, which is directly related to the power spectral density (PSD) slope. For a DRW process we expect $\gamma=0.5$.

\subsection{Sample filtering by light curves properties}\label{samp_lc}

\begin{figure}
\begin{center}
\includegraphics[scale=0.5]{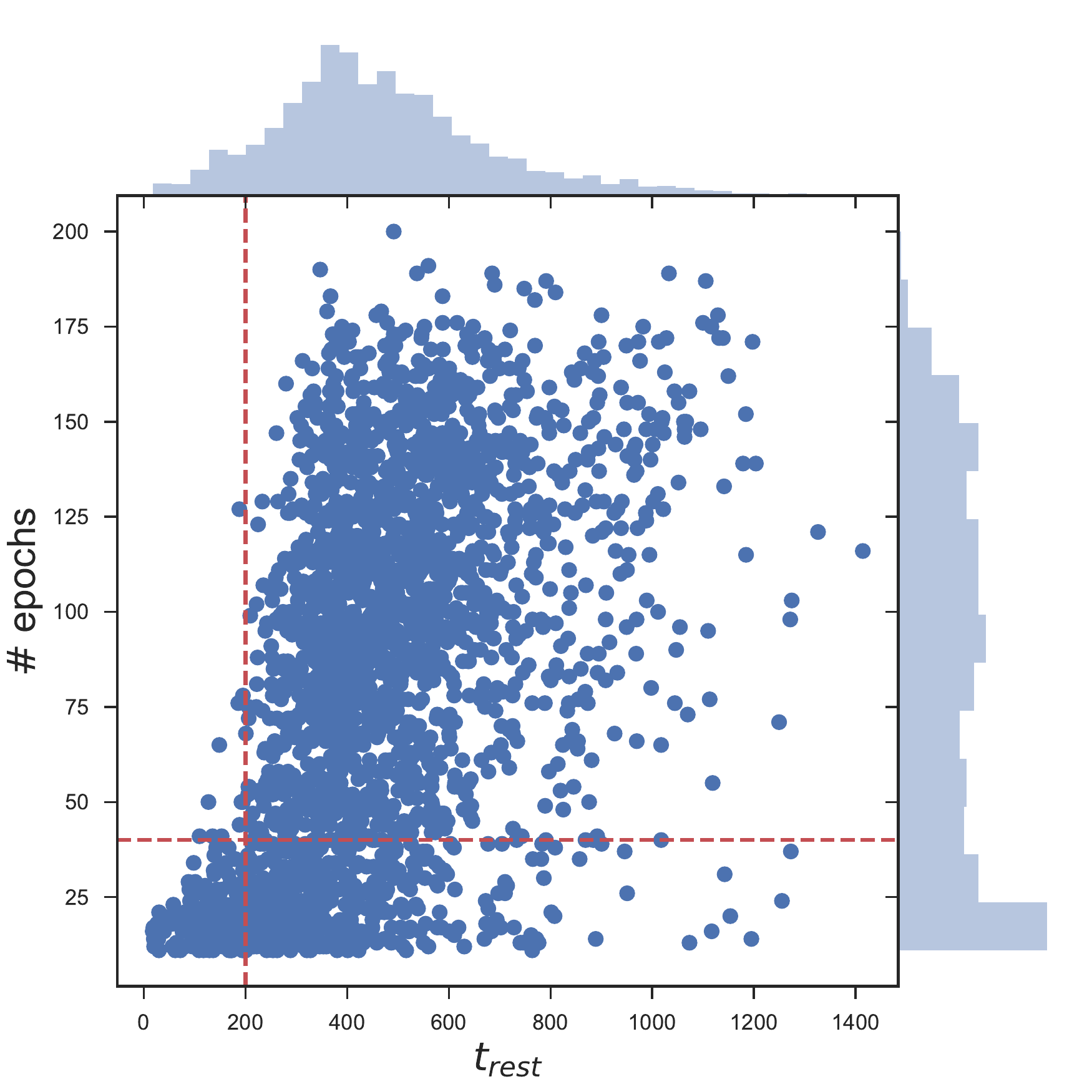}
\caption{Number of epochs vs. rest frame time length of the 2345 light curves with SDSS spectra. The red vertical dashed line shows the position where $t_{\text{rest}}\equiv 200$ days, and the red horizontal dashed line shows the position where $\#\text{epochs} \equiv 40$. \label{figure:epochs_time}}
\end{center}
\end{figure}

Since we want to study individual light curves, we have to consider only those sources with sampling dense enough to get statistically significant variability features. Figure \ref{figure:epochs_time} shows the distribution of the number of epochs ($\#\text{epochs}$) and rest frame time length of our 2345 light curves with SDSS spectra. Following a similar approach than S17, we selected for our analysis those light curves with $t_{\text{rest}}\geqslant 200$ days, and in order to ensure a high number of epochs, we also selected those light curves with $\#\text{epochs} \geqslant 40$. In the figure we can see that most of our sources satisfy these conditions. After we filter our sample by the number of epochs and the length of the light curve, we ended with 1751 sources. Hereafter, we refer to it as the `well-sampled'  sub-sample.

\subsection{Biases of the variability features}\label{var_bias}

\begin{figure}
\begin{center}
\includegraphics[scale=0.5]{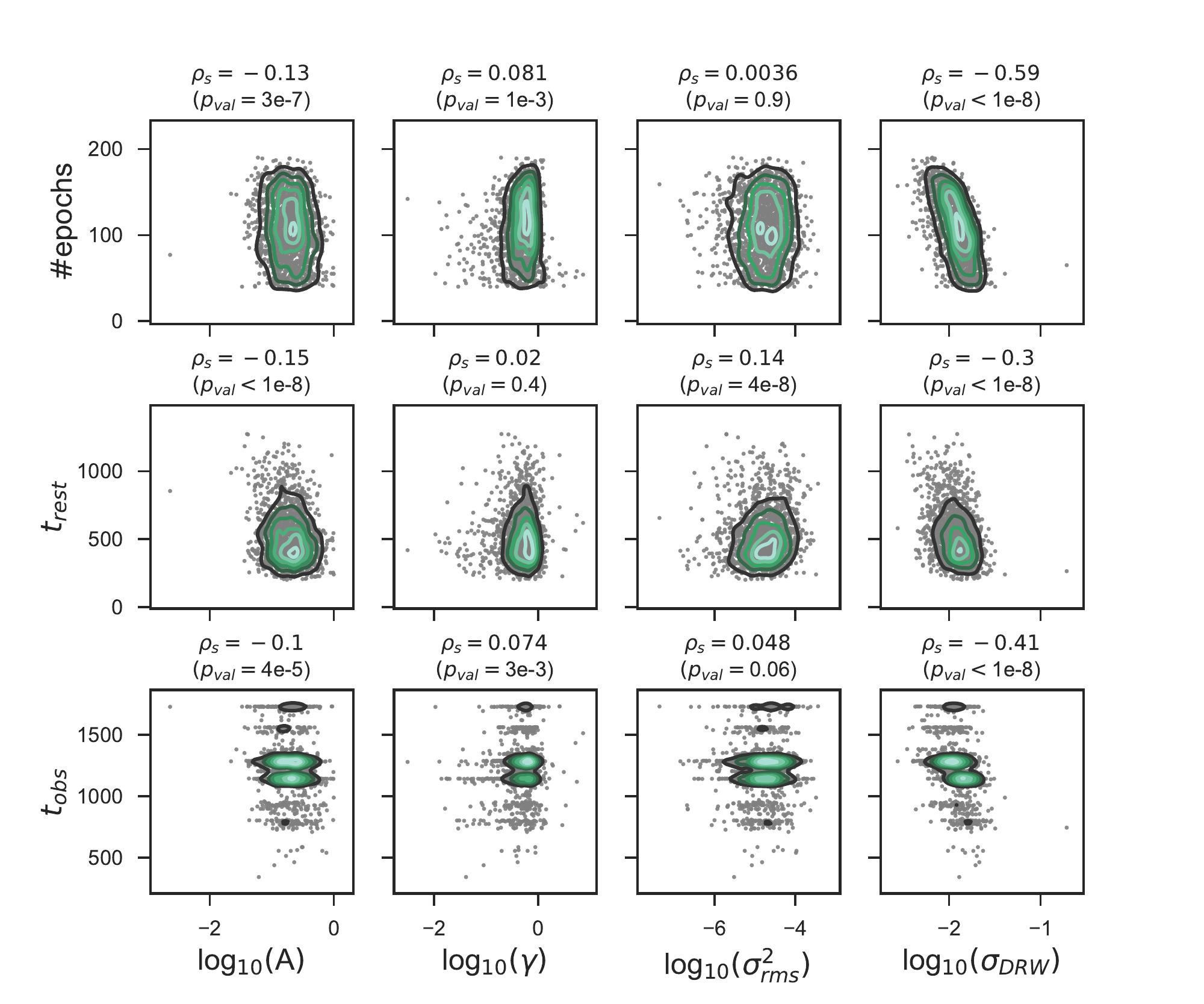}
\caption{Variability features vs. light curve properties. The Spearman's rank correlation coefficient is shown as reference for every pair of variables. \label{figure:varfeat_lcprop}}
\end{center}
\end{figure}

Figure \ref{figure:varfeat_lcprop} shows the different measured variability features versus the light curve properties (i.e. the number of epochs or the length in days). We can see that the SF parameters $\gamma$ and $A$  are practically unaffected by the light curve length and the number of epochs. The excess variance $\sigma_{rms}$ is also unaffected by the light curve sampling. On the other hand, $\sigma_{DRW}$ is affected by the length of the light curve and strongly affected by the number of epochs. This is quantified by the Spearman's rank coefficient which gives values of -0.59 ($p_{val}<$1e-8), -0.3 ($p_{val}<$1e-8) and -0.41 ($p_{val}<$1e-8) for the correlations of $\sigma_{DRW}$ and \#epochs, $t_{rest}$, and $t_{obs}$, respectively. We decided to use $A$ and $\gamma$ as the main features for our analysis, and use $\sigma_{DRW}$ and $\sigma_{rms}$ as references.

\subsection{Variability of simulated light curves}\label{sim_SF}

In this work, we used the parameters of the structure function as the main variability features, therefore it is important to understand how these parameters respond to different factors.

\begin{figure}
\begin{center}
\includegraphics[scale=0.5]{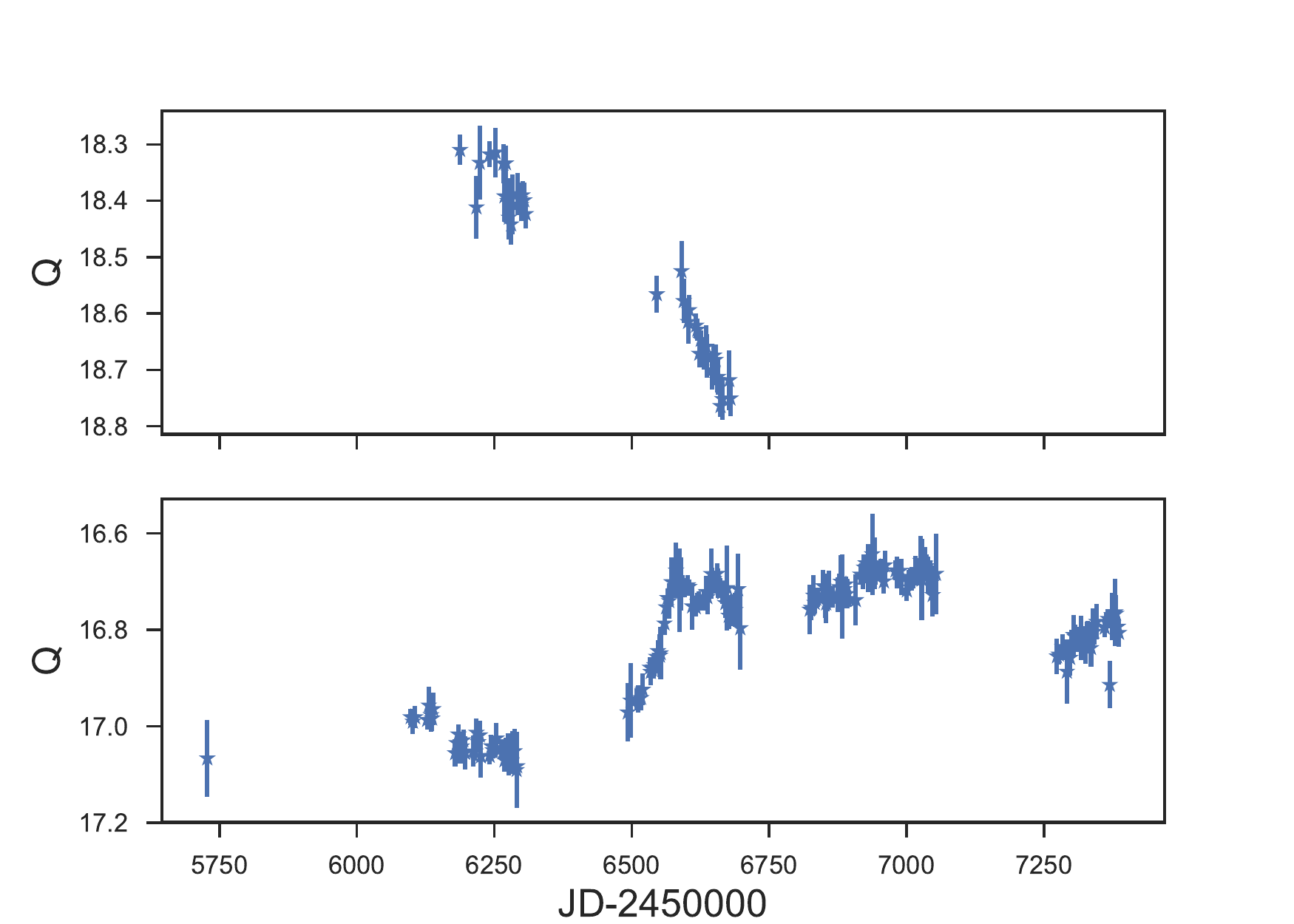}
\caption{Light curves with different number of epochs and length, but similar cadence. \label{figure:lc_plot}}
\end{center}
\end{figure}

In order to understand how the sampling of a given light curve affects the measurement of its SF parameters, we simulated artificial light curves, following a similar approach than S17 (see their sections 5.3 and 5.4). We simulated light curves using the same sampling of the light curves shown in Figure \ref{figure:lc_plot}. The light curve at the top of the figure (short light curve), has a length of 492 days and 39 observing epochs. The light curve at the bottom (long light curve) has 175 observing epochs and a length of 1659 days.

\begin{figure}
\begin{center}
\includegraphics[scale=0.35]{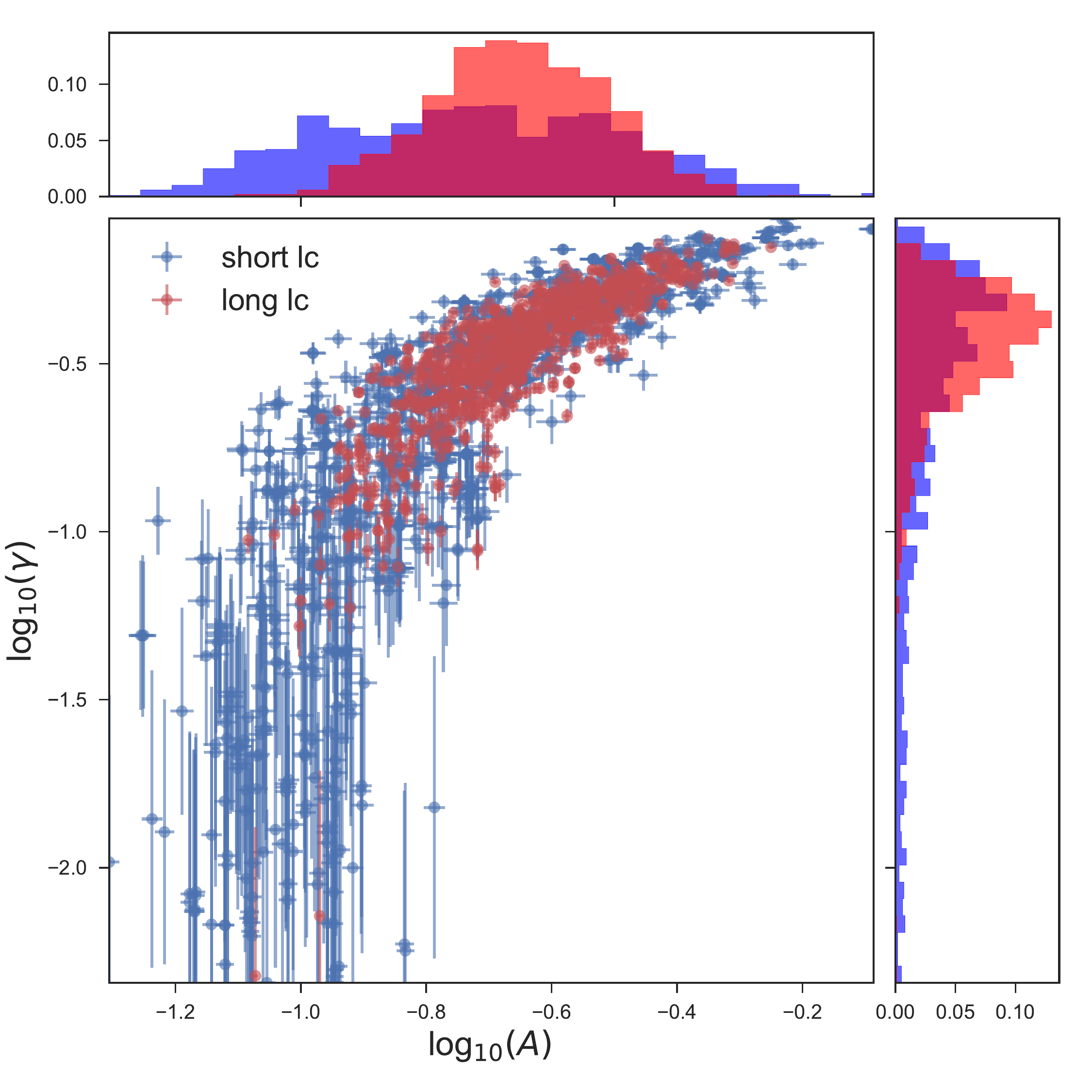}
\caption{SF parameters measured for 1000 light curves simulated from a DRW process with $\tau=300$ days and $\text{SF}_{\text{inf}}=0.2$ mag, with short and long samplings. Along the axes we show the histograms of every parameter. \label{figure:SF_drw_sim}}
\end{center}
\end{figure}

We simulated light curves from a DRW process with $\tau=300$ days and $\text{SF}_{\text{inf}}=0.2$ mag. For each long and short light curves, we simulated 1000 light curves. Figure \ref{figure:SF_drw_sim} shows the results of the SF parameters measured for the short and long simulated light curves. First of all, we can see that the distributions of the parameters measured for the short light curves have a larger dispersion than the values measured for the long light curves. For the case of the short light curves, the median, mean and standard deviation of $A$ are 0.19, 0.22 and 0.12 respectively, and for $\gamma$ are 0.3, 0.32 and 0.21 respectively. For the case of the long light curves, the median, mean and standard deviation of $A$ are 0.22, 0.23 and 0.07 respectively, and for $\gamma$ are 0.37, 0.37 and 0.14 respectively. From Figure \ref{figure:SF_drw_sim}, we can also see that there is a correlation between the measured parameters. For the case of the short light curves the Spearman's rank coefficient is $\rho_s= 0.91$ ($p_{val}<$1e-8), and for the long light curves we have $\rho_s= 0.87$ ($p_{val}<$1e-8). 

We tested whether other definitions of the SF show the same behavior, using the definition of \cite{Kozlowski16a} and the ``traditional'' definition of \cite{Schmidt10}. The results were consistent with what was found for the Bayesian method of \cite{Schmidt10}. In addition, we simulated artificial light curves with a power-law PSD (assuming different values for the exponent), following the same approach of S17, and we found similar results. 

We also tested whether longer light curves can solve the degeneracy between the SF parameters. We simulated 1000 light curves from a DRW process with $\tau=300$ days and $\text{SF}_{\text{inf}}=0.2$ mag, with 7000 days and 700 epochs, and with a similar cadence than the long light curve of figure \ref{figure:lc_plot}. The Spearman's rank coefficient of $A$ versus $\gamma$ is $\rho_s= 0.88$ ($p_{val}<$1e-8), therefore, the parameters are still correlated. The median, mean and standard deviation of $A$ are 0.24, 0.24 and 0.04 respectively, and for $\gamma$ are 0.39, 0.39 and 0.07 respectively. In order to see whether the previous results are produced by the gaps in the data, we simulated 1000 light curves from a DRW process with $\tau=300$ days and $\text{SF}_{\text{inf}}=0.2$ mag, with 7000 days of length and with observations every 10 days, obtaining similar results for the correlation between $A$ and $\gamma$.

We simulated white noise light curves, with a standard deviation of 0.2 mag, in order to test whether the correlation between the SF parameters is also present for light curves with a constant PSD. The Spearman's rank coefficient for the short light curves is $\rho_s= 0.23$ ($p_{val}<$1e-8), and for the long light curves is $\rho_s= 0.23$ ($p_{val}<$1e-8). In this case, we also see a broader distribution of the parameters for the short light curves. For the case of the short light curves, the median, mean and standard deviation of $A$ are 0.29, 0.29 and 0.04 respectively, and for $\gamma$ are 0.02, 0.02 and 0.02 respectively. For the case of the long light curves, the median, mean and standard deviation of $A$ are 0.28, 0.28 and 0.02 respectively, and for $\gamma$ are 0.006, 0.007 and 0.005 respectively. We also simulated white noise light curves, with a standard deviation of 0.02 mag and we found similar results.

\begin{figure}
\begin{center}
\includegraphics[scale=0.4]{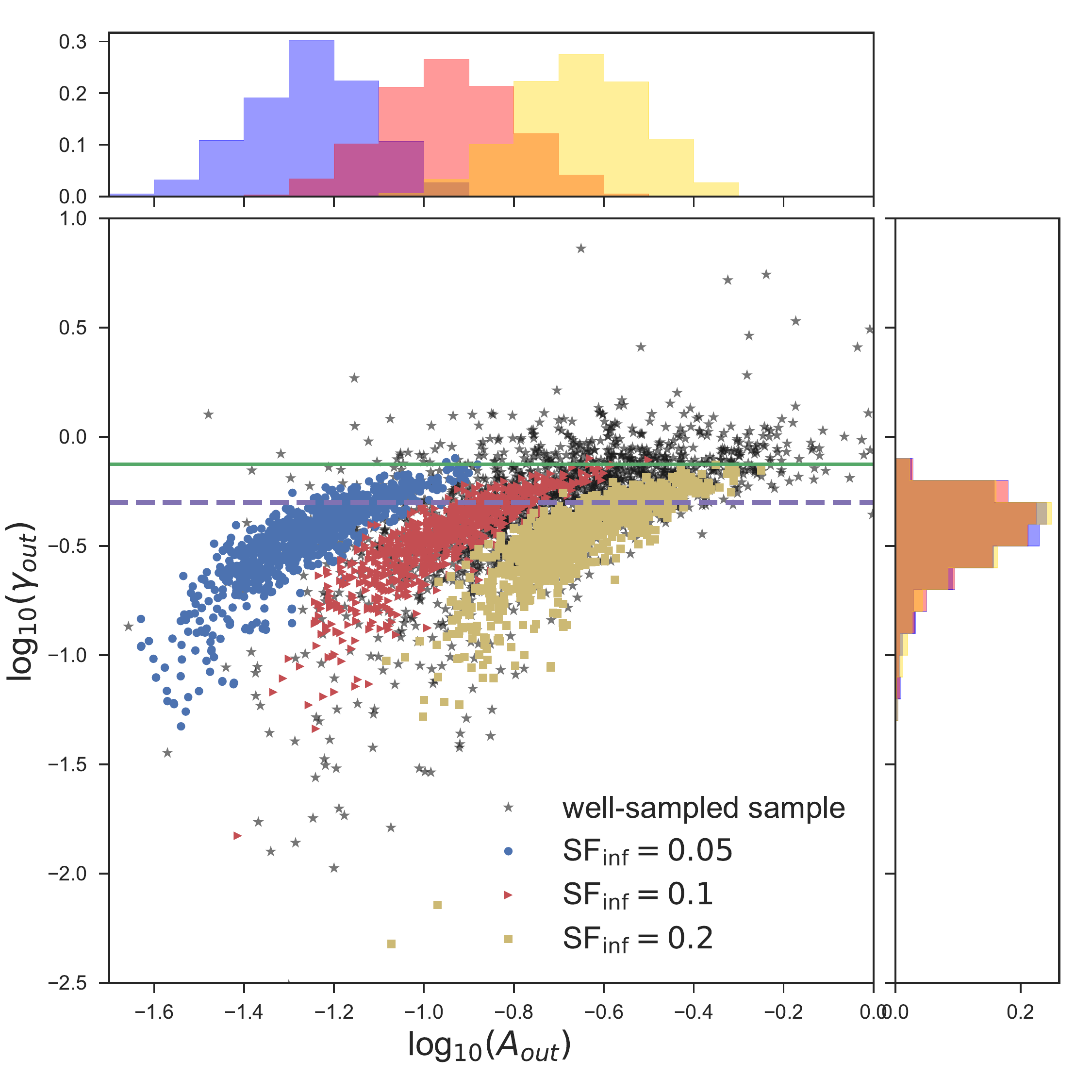}
\caption{SF parameters measured for light curves simulated from a DRW process with $\tau=300$ days and different values of $\text{SF}_{\text{inf}}$ (with the long sampling). The blue circles correspond to light curves simulated with $\text{SF}_{\text{inf}}=0.05$, the red triangles correspond to light curges simulated with $\text{SF}_{\text{inf}}$ =0.1, and the yellow squares correspond to light curves simulated with $\text{SF}_{\text{inf}}=0.2$. Along the axes we show the histograms of every parameter for the different $\text{SF}_{\text{inf}}$. The green solid line shows the position where $\gamma=0.75$, and the magenta dashed line show the position where $\gamma=0.5$, the expected value for a DRW process. We show with black stars the measurments done for the variable sources of the well-sampled sub-sample.\label{figure:sim_A_vs_g}}
\end{center}
\end{figure}

In order to test whether the distribution of $A$ and $\gamma$ values measured for the real light curves is simply produced by this degeneracy and scatter, we compare the regions in the $A-\gamma$ plane covered by the real and simulated light curves. We simulated DRW light curves with $\tau=300$ days and with different values of $\text{SF}_{\text{inf}}$. For every amplitude we simulated 1000 light curves, with the same sampling of the long light curve in Figure \ref{figure:lc_plot}. Figure \ref{figure:sim_A_vs_g} shows the measured values of $A$ and $\gamma$ for three different values of $\text{SF}_{\text{inf}}$ (0.05, 0.1, and 0.2). In the figure we can see that there is no change in the distribution of $\gamma$ for different values of $\text{SF}_{\text{inf}}$. We can also see that independently of the value of $\text{SF}_{\text{inf}}$, the measured values of $\gamma$ range from 0.0 to 0.75. Therefore, if we measure a value of $\gamma$ between this range for a given light curve, we cannot discard a DRW process as the best model to describe the variability. However, the distribution of gamma values is significantly different for the simulated and real light curves, so as a  population the AGN light curves are not well represented by a DRW model, at least because it incorporates many outliers

As can be seen in Figure \ref{figure:sim_A_vs_g}, no single value of the intrinsic amplitude $\text{SF}_{\text{inf}}$ reproduces the entire parameter space, from where we conclude that the measured values of the amplitude correlate on average with the intrinsic amplitude.

\begin{figure}
\begin{center}
\includegraphics[scale=0.5]{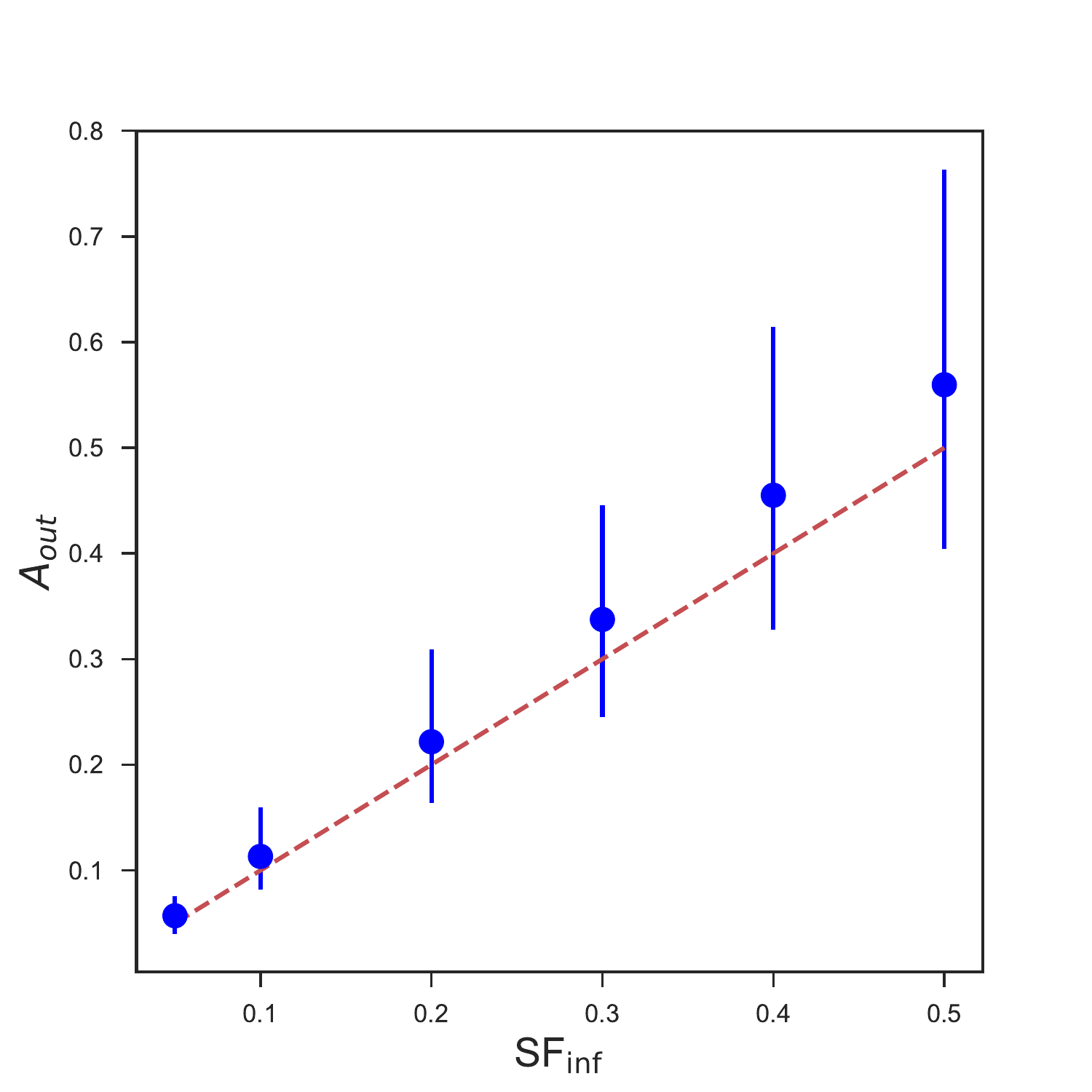}
\caption{$A$ measured for light curves simulated from a DRW process with $\tau=300$ days and different values of $\text{SF}_{\text{inf}}$ (with the long sampling). The circles correspond to the median value measured, and the error bars correspond to the 15.9 and 84.1 percentiles. The red dashed line shows the expected value of $A$ (1:1 relation). \label{figure:sim_Aobs_vs_Ain}}
\end{center}
\end{figure}

Figure \ref{figure:sim_Aobs_vs_Ain} shows the distribution of measured values of $A$ for different input $\text{SF}_{\text{inf}}$. We can see that we can recover the input amplitude value, but with larger dispersions for larger values of $\text{SF}_{\text{inf}}$. Therefore, we can say that the amplitude measured with the structure function is a good estimator of the intrinsic variability amplitude, albeit with significant scatter due to the $A - \gamma$ degeneracy.

\begin{figure}
\begin{center}
\includegraphics[scale=0.5]{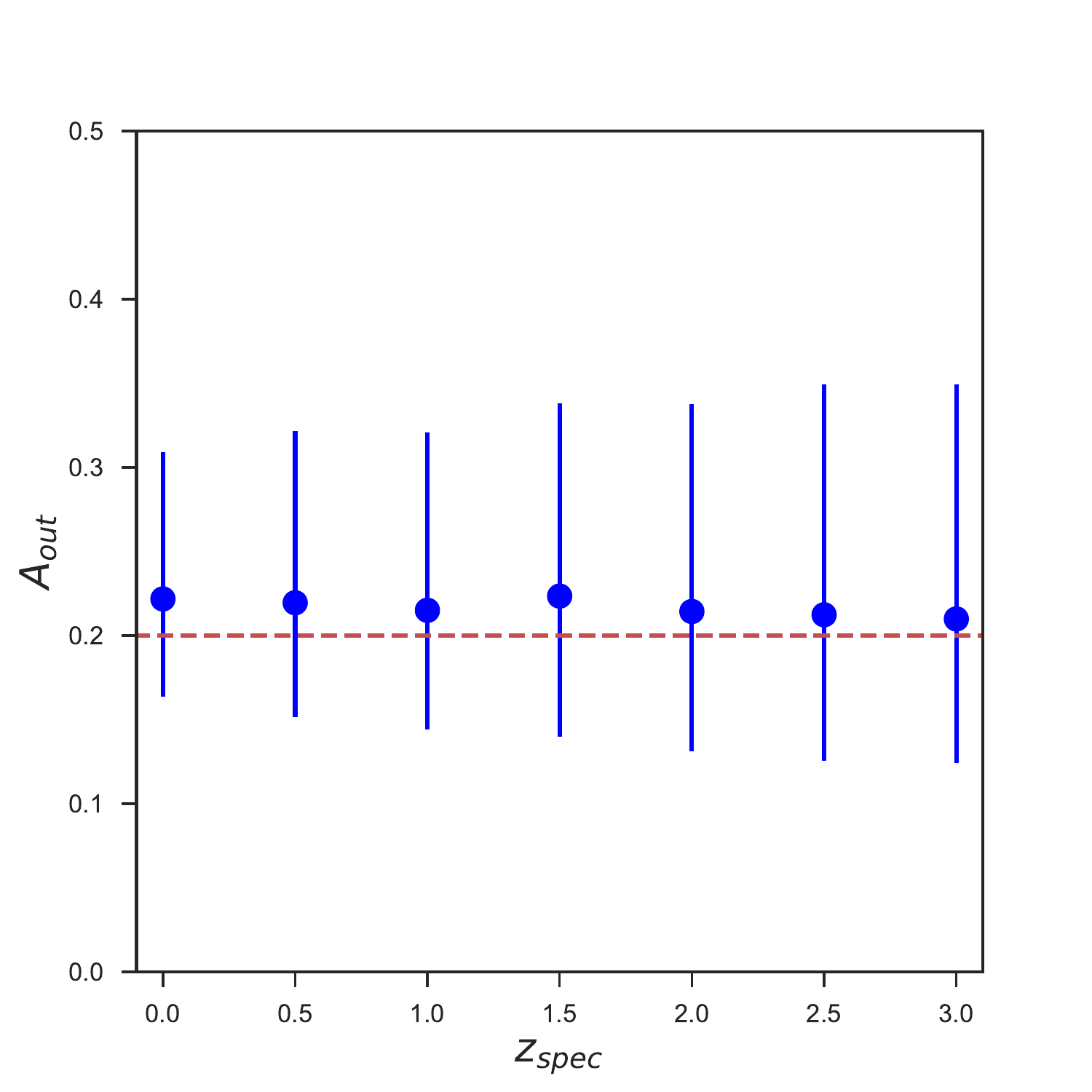}
\caption{$A$ measured for light curves simulated from a DRW process  with $\tau=300$ days and $\text{SF}_{\text{inf}}=0.2$ mag The circles correspond to the median value measured, and the error bars correspond to the 15.9 and 84.1 percentiles. The red dashed line shows the expected value $A=0.2$. \label{figure:sim_Aobs_vs_z}}
\end{center}
\end{figure}

\begin{figure}
\begin{center}
\includegraphics[scale=0.5]{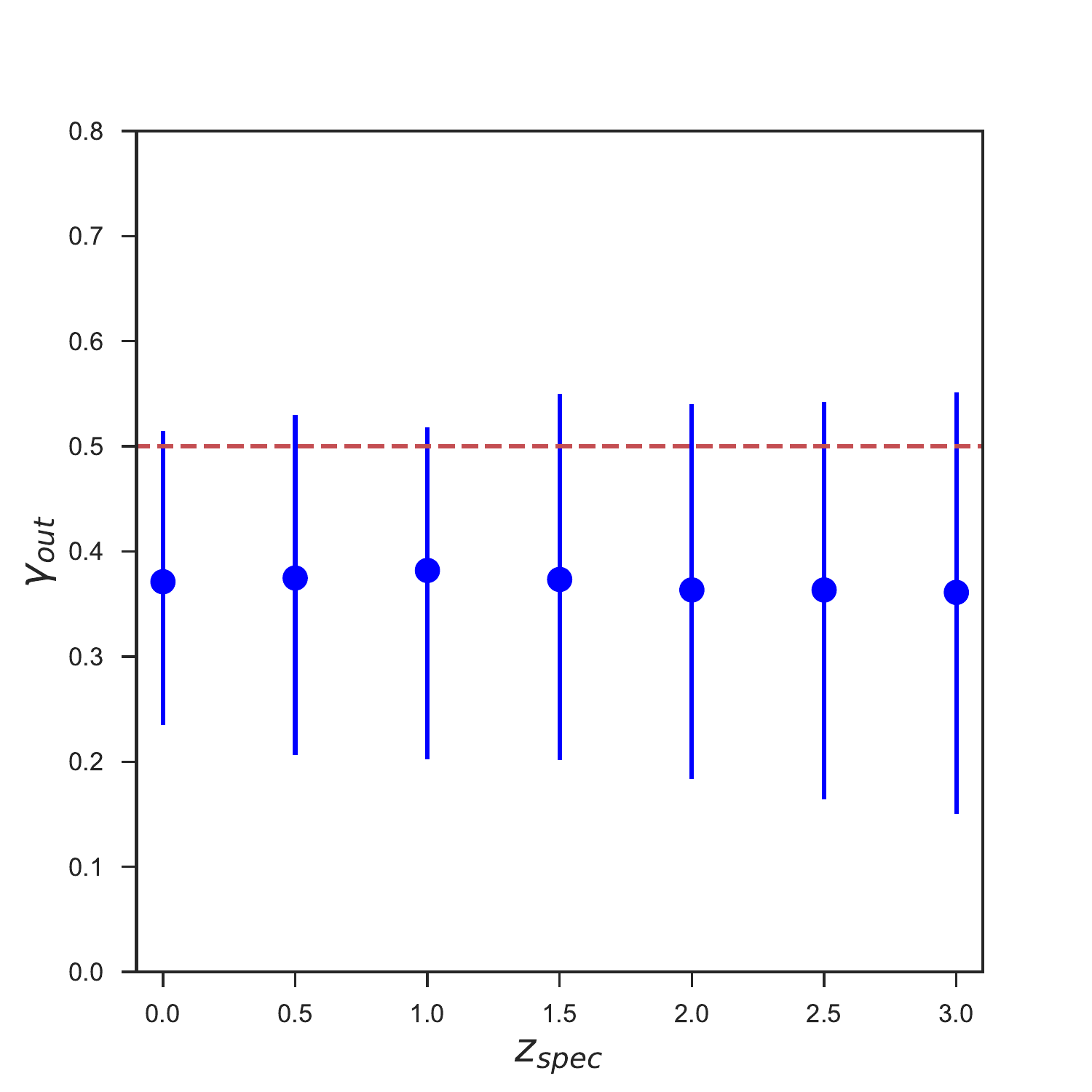}
\caption{$\gamma$ measured for light curves simulated from a DRW process  with $\tau=300$ days and $\text{SF}_{\text{inf}}=0.2$ mag The circles correspond to the median value measured, and the error bars correspond to the 15.9 and 84.1 percentiles. The red dashed line shows the expected value $\gamma=0.5$.  \label{figure:sim_gobs_vs_z}}
\end{center}
\end{figure}

We also tested the effects of the source redshift in the measured parameters, since sources at higher redshifts have shorter rest frame light curves for a given observed light curve ($t_{rest}=t_{obs}/(1+z_{spec})$). We simulated DRW light curves with $\tau=300$ days and $\text{SF}_{\text{inf}}=0.2$. In order to account for the redshift of the source, we use the sampling of the long light curve of Figure \ref{figure:lc_plot}, dividing the time by $(1+z_{spec})$, for different values of $z_{spec}$. We simulated 1000 light curves per every value of $z_{spec}$. Figures \ref{figure:sim_Aobs_vs_z} and \ref{figure:sim_gobs_vs_z} show the results for $A$ and $\gamma$ respectively. We can see that both values do not change with $z_{spec}$, but the dispersion of the measured values increase a little bit with redshift. We can also see that the measured value of $A$ is close to the input value (red dashed line in the figure), but the measured value of $\gamma$ is below the expected value. This is consistent with the findings of S17 (see their Figure 7).

From all these results, we can conclude that the $A - \gamma$ degeneracy is much lower for the case of light curves with a constant PSD. For the case of stochastic light curves, the broad distribution of the measured parameters is produced by the fact that light curves with a few months or years of coverage are not a well representation of the general behavior of variability with decorrelation time-scales of months or years, or with power-law PSD, and by the $A-\gamma$ degeneracy. The correlation between the measured SF parameters is present independently on the SF definition used. We can reduce the effects of this degeneracy by using light curves with several years of length. Besides, we can conclude that the values of the amplitude of the variability obtained from the SF, are a good estimation of the real amplitude, independently of the redshift of the source. On the other hand, for the case of $\gamma$, we must consider that if we measure a value between 0.0 and 0.75 we cannot discard a DRW process as the best model to describe the variability.

\subsection{Variability of the QUEST light curves}\label{var_quest}

\begin{figure}
\begin{center}
\includegraphics[scale=0.38]{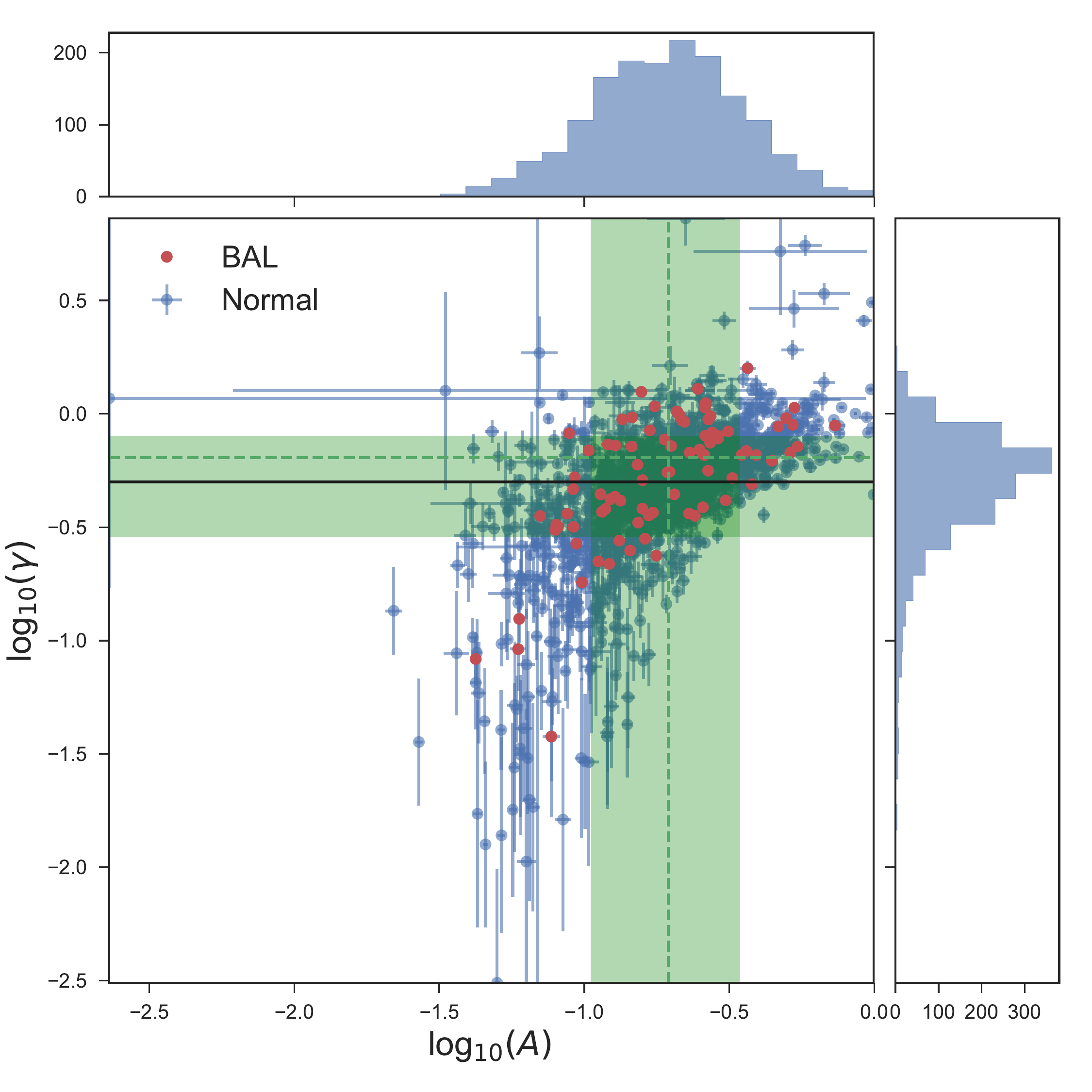}
\caption{Distribution of the SF parameters $A$ and $\gamma$ for the variable and well sampled sources. Along the axes we show the histograms of every parameter. We plot in red the sources classified as BAL QSO, and in blue rest of the sample. The black solid line shows the expected value of $\gamma$ for a DRW process. The green dashed lines show the median of the parameters. The green shaded regions show the 15.9 to 84.1 percentile range.  \label{figure:SF_feat}}
\end{center}
\end{figure}

Following the criteria that sources with $P_{var} \geq 0.95$ and $(\sigma^2_{rms}-err(\sigma^2_{rms}))>0$ are classified as variable, 1579 of the well-sampled light curves are variable, which corresponds to the $90.2 \%$ of the sample. Figure \ref{figure:SF_feat} shows the distribution of the SF parameters, for the variable sources of the well-sampled sub-sample. The weighted average of the parameters are $\bar{A}=0.21 \pm 0.12$ and $\bar{\gamma}=0.64 \pm 0.22$. The median, and the percentiles 15.9 and 84.1 of the measured values of $A$ are 0.19, 0.11, and 0.34 respectively. The median, and the percentiles 15.9 and 84.1 of $\gamma$ are 0.53, 0.39, and 0.80 respectively. The measured values of $A$ are consistent with previous findings for optical variability (e.g. \citealt{Schmidt10,MacLeod10,Cartier15,Suberlak17,Rakshit17}). From Figure \ref{figure:sim_A_vs_g} we can see that our measurements are consistent with amplitudes ranging between 0.05 and 0.2, however there are some sources with larger amplitudes.

For a DRW process, the expected value of $\gamma$ is 0.5, however, in section \ref{sim_SF}, we showed that the Bayesian method tend to underestimate the value of $\gamma$ (see Figures \ref{figure:sim_A_vs_g} and \ref{figure:sim_gobs_vs_z}). From Figure \ref{figure:sim_A_vs_g}, we can see that the distribution of $\gamma$ measured for the well-sampled sub-sample (black stars) differs from the distribution of $\gamma$ obtained for light curves simulated from a DRW process with different amplitudes. Therefore we cannot say that a DRW process can explain the general behavior of the variability in our sample.

\begin{figure*}
\begin{center}
\includegraphics[scale=0.35]{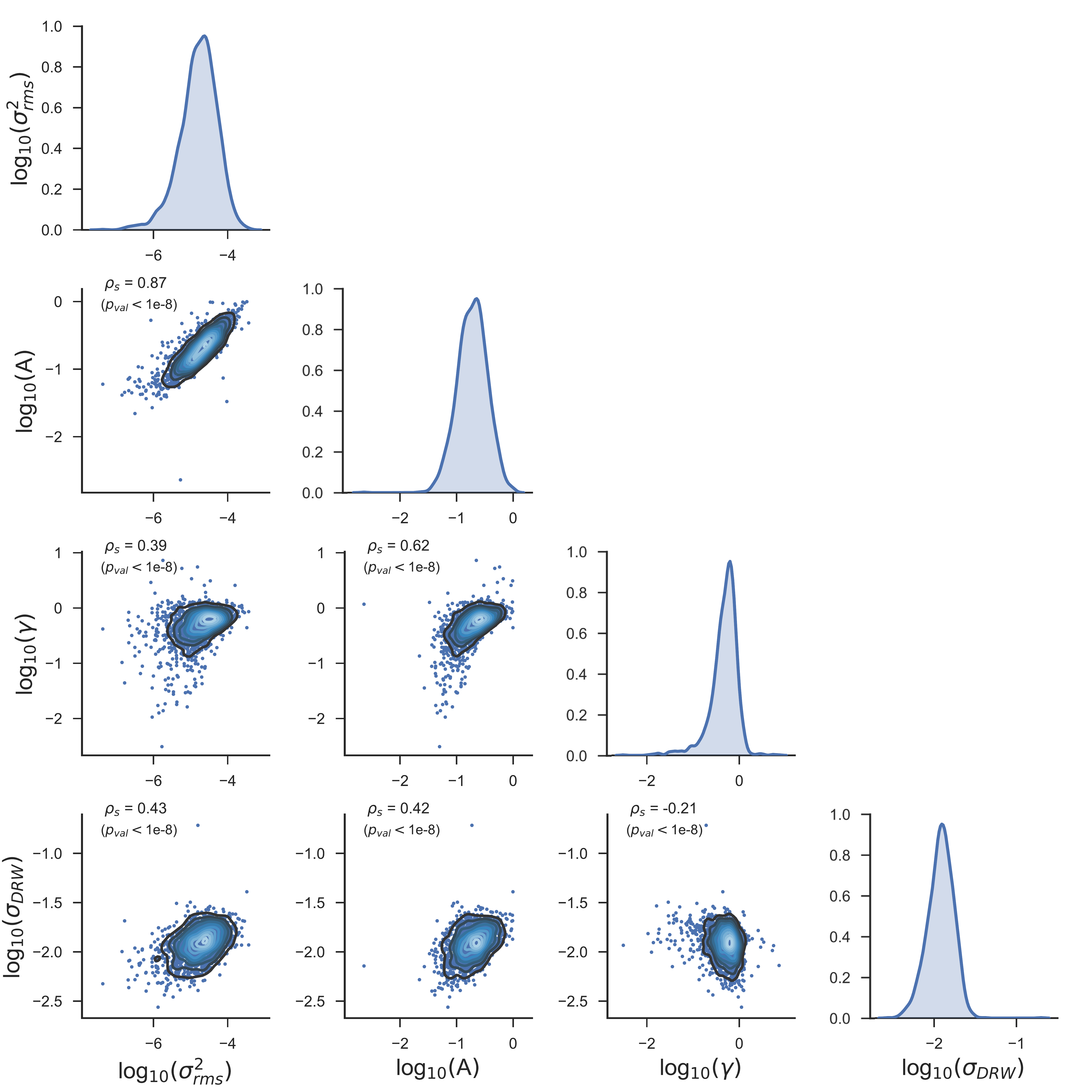}
\caption{Correlations of the variable features for all the variable and well sampled sources. The diagonal shows the individual distributions. As a reference we provide the Spearman's rank correlation coefficient for every pair of variables.\label{figure:var_feat}}
\end{center}
\end{figure*}

Figure \ref{figure:var_feat} shows the correlations of the variable features (2D distributions) and their individual distributions (1D distributions) for all the sources classified as variable in the well-sampled sub-sample. Our results for $\sigma_{DRW}$ are consistent with \cite{Kelly09}. Our measurement of $\sigma^2_{rms}$ can be characterized by a mean and standard deviation of $(2.8\pm3.3)\times 10^{-5}$, which is smaller than the values reported by previous analyses (e.g. \citealt{Cartier15,Simm16}). When we compare the results for $\sigma^2_{rms}$, we must consider that it depends on the length of the light curve, since this parameter consider the total variance of the light curve, and also it depends on the photometric errors, since with low photonetric errors we can detect lower variability amplitudes. It is also important to consider the differences in the deffinition of the excess variance used by different authors. For example, \cite{Cartier15} used a non-normalized version of the excess variance for optical light curves measured in magnitudes and with a shorter coverage in time compared to our light curves, but with the same photometric errors, since they used QUEST-La Silla data. If we correct their measurements by the mean magnitude, we obtain consistent results. \cite{Simm16} used optical light curves measured in flux and with different coverage in time, which introduces differences with our results.

We can also see from Figure \ref{figure:var_feat} that the strongest correlation is shown between $A$ and $\sigma^2_{rms}$. This can be produced by the sampling of the light curve, since several light curves has lengths in rest frame close to 1 year (see Figure \ref{figure:epochs_time}). 

From Figures \ref{figure:SF_feat} and \ref{figure:var_feat}, we can see that the SF parameters are correlated. The  Spearman's rank correlation coefficient for $A$ versus $\gamma$ is $\rho_s=0.62$ ($p_{val}<$1e-8). In section \ref{sim_SF} we showed that this correlation is a product of the degeneracy in the SF parameters, produced by the stochastic nature of the light curves, the sampling of the light curves, and the structure function method by itself. Therefore it does not have any physical implication.

\section{Spectral analysis}\label{spec_analysis}

\subsection{Host galaxy subtraction}\label{hot_gal_sub}

For low redshift sources ($z_{spec}\leq 0.8$) we can have a significant degree of host galaxy contamination in the optical SDSS spectra, depending on the brightness of the nucleus. We follow the simple procedure to substract the galactic continuum of \cite{Greene05} and \cite{Kim06}, where the stellar continuum is modelled using the scaled spectrum of a K giant star. In order to know how much starlight must be subtracted from the spectra, we use the equivalent width (EW) of the Ca II K absorption line ($\lambda = 3934$). 

We isolated the AGN component only for those objects with  $z_{spec}\leq0.8$ (349 sources), since for sources with $z_{spec}>0.8$ we can ensure the presence of the Mg II line in the SDSS spectra, and the H$_{\beta}$ line would be located in the edges of the spectra. For 304 of these 349 sources, the quality of the spectra was good enough to obtain the AGN and host galaxy components (i.e. with $S/N$(Ca II K)$>10$).

\subsection{Spectral fitting and measurement of physical properties}\label{spec_fit}

We used the procedure proposed by MR16 to estimate the black hole masses ($M_{\text{BH}}$), the luminosity at 5100{\AA} ($\text{L}_{5100}$), the accretion rate ($\dot{M}$), and the Eddington ratio ($L/L_{\text{Edd}}$) for our AGN. MR16 proposed new calibrations for the measurement of these physical properties from single-epoch spectrum, by fitting the $\text{H}_{\alpha}$, $\text{H}_{\beta}$, Mg II $\lambda 2798$ and C IV $\lambda 1549$ lines. Their method relies on the assumption of virialized BLR kinematics and consider the FWHM of the line as a proxy to the virial velocity of the gas in the BLR ($V_{\text{BLR}}$). Additionally, the continuum luminosity in the proximity of the emission line is used to estimate the BLR radius by means of the empirical luminosity-radius relationship derived from several reverberation mapping experiments (e.g. \citealt{Bentz13}). Their method model the broad emission lines with two broad Gaussian components, and for the case of doublet lines (Mg II and C IV) they use two additional Gaussians, which are separated by the theoretical wavelength doublet separation, and are forced to have the same profiles and intensity of the other two Gaussians, which is valid for optically thick BLR clouds. For the case of $\text{H}_{\alpha}$ and $\text{H}_{\beta}$ their method also includes a third narrow line component, modelled with a single Gaussian, which account for the narrow line emitting region. To model the continuum emission of the AGN, they followed the local approach described in MR16 which consists of fitting a single power-law limited within two narrow pseudo-continuum windows around the emission line. For the case of the $\text{H}_{\beta}$ and Mg II lines, their method also includes the modelling of an iron pseudo-continuum, using an iron template, which originates from a large number of blended features of Fe II and Fe III (for further details see MR16).

The black hole mass is calculated as $M_{\text{BH}}=fG^{-1}R_{\text{BLR}}V^2_{\text{BLR}}=K(\lambda L_{\lambda})^{\alpha}\text{FWHM}^2$. The values of $K$ and $\alpha$ for every line can be found in Table 7 of MR16. The accretion rate is estimated from equation 1 of \cite{Netzer14}, as follows $$ 4\pi D_{\text{L}}^2 F_{\nu} = f(\theta) [M_{8}\dot{M}]^{2/3} \left[ \frac{\lambda}{5100{\AA}} \right]^{-1/3} \text{erg } \text{s}^{-1} \text{Hz}^{-1}, $$ where $M_{8}$ corresponds to $M_{\text{BH}}$ measured in units of $10^8\text{M}_{\odot}$, $\dot{M}$ is the accretion rate in units of $M_{\odot}/\text{year}$, $D_{\text{L}}$ is the luminosity distance, and $f(\theta)$ is the inclination-dependent term, that describes the orientation of the accretion disk to the line of sight.

We used the calibrations derived by MR16 to estimate $\text{L}_{5100}$ from $\text{L}_{6200}$, $\text{L}_{3000}$ and $\text{L}_{1450}$ (see their Table 5). From this, we estimated Eddington ratio as $$L/L_{\text{Edd}}=\frac{C_{\text{BOL}}\text{L}_{5100}}{1.5\times 10^{38}(M_{\text{BH}}/M_{\odot})},$$ where $C_{\text{BOL}}$ is the bolometric correction. In this analysis we adopted $C_{\text{BOL}}=9.26$ (see \citealt{Shen08,MacLeod10} and references therein).

\cite{MejiaRestrepo18a} proposed new corrections for the estimation of $M_{\text{BH}}$, which intend to account for the effect of the unknown distribution of the gas clouds in the BLR. They suggest that the virial correcting factor is inversely proportional to the width of the broad emission line used to compute $M_{\text{BH}}$. This can be explained either by line of sight inclination effect on a planar BLR or by radiation pressure effects on the BLR gas distribution \citep{Kollatschny13}. The corrected black hole mass is calculated as $M_{\text{BH}}^C=fM_{\text{BH}}$, with $f=(\text{FWHM}_{\text{obs}}(\text{line}) / \text{FWHM}_{\text{obs}}^0)^{\beta}$, where $\text{FWHM}_{\text{obs}}(\text{line})$ is the FWHM of the emission line.  $\text{FWHM}_{\text{obs}}^0$ and $\beta$ are parameters calculated per every line (see Table 1 in \citealt{MejiaRestrepo18a}). We computed $M_{\text{BH}}^C$ for our sample, and used it to calculate the corrected Eddington ratio $(L/L_{\text{Edd}})^C$. In the next sections we will compare our results when using both the original and the corrected black hole masses.

\subsection{Spectral properties of the selected SDSS spectra}\label{spec_prop}

We have a total of 2345 sources with SDSS spectra, however not all of them have a signal to noise ($S/N$) high enough to allow the fitting of the emission lines. In our analysis we only consider those spectra with a mean $S/N$ per pixel, in the continuum region around the emission line of interest, larger or equal to 10. We also exclude from our analysis those sources classified as Broad Line Absorption QSO (BAL) in the catalogs of \cite{Shen11}, \cite{Paris17a} and \cite{Paris17b}, with strong absorption lines in the region of the emission line under analysis. After we eliminated sources from our sample with low $S/N$ and classified as BAL QSO, we end with 102 sources having $\text{H}_{\alpha}$ inside the SDSS wavelength coverage, 304 sources with $\text{H}_{\beta}$, 1561 sources with Mg II, and 801 sources with C IV. Most of our sources have more than one line available.

We fit the $\text{H}_{\alpha}$ and $\text{H}_{\beta}$ lines for sources with $z_{spec}\leq 0.8$). For 81 sources we could obtain a fit of $\text{H}_{\alpha}$ and for sources 224 we obtained a fit of $\text{H}_{\beta}$. Besides, we obtained a fit of Mg II for 1487 objects, and a fit of C IV for 718. 

After the line fitting, we calculated $M_{\text{BH}}$, $\text{L}_{5100}$, $\dot{M}$, and $L/L_{\text{Edd}}$ using the equations given in section \ref{spec_fit}. We considered only those line fits where the height to noise ($H/N$) of the line is $H/N\geq5$, with the height defined as the distance between the peak of the line fitted and the continuum. This is done to avoid the fit of fake lines when the broad lines are weak or are not present. Since for some objects we have more than one line available, we decided to estimate the final $M_{\text{BH}}$ as the weighted average of the measured $M_{\text{BH}}$ for the different lines, with the exception of the C IV line. It is well known that the C IV line width is not a good estimator of $V_{\text{BLR}}$, and therefore the measurements done using this line must be taken carefully (see MR16 and references therein). Therefore, whenever a source has C IV and other lines available, we excluded C IV from the estimation of $M_{\text{BH}}$, and we only consider the results of C IV when there is no other line available. In the next sections, the analyses are done with and without the results of the C IV fitting.

\begin{figure*}
\begin{center}
\includegraphics[scale=0.38]{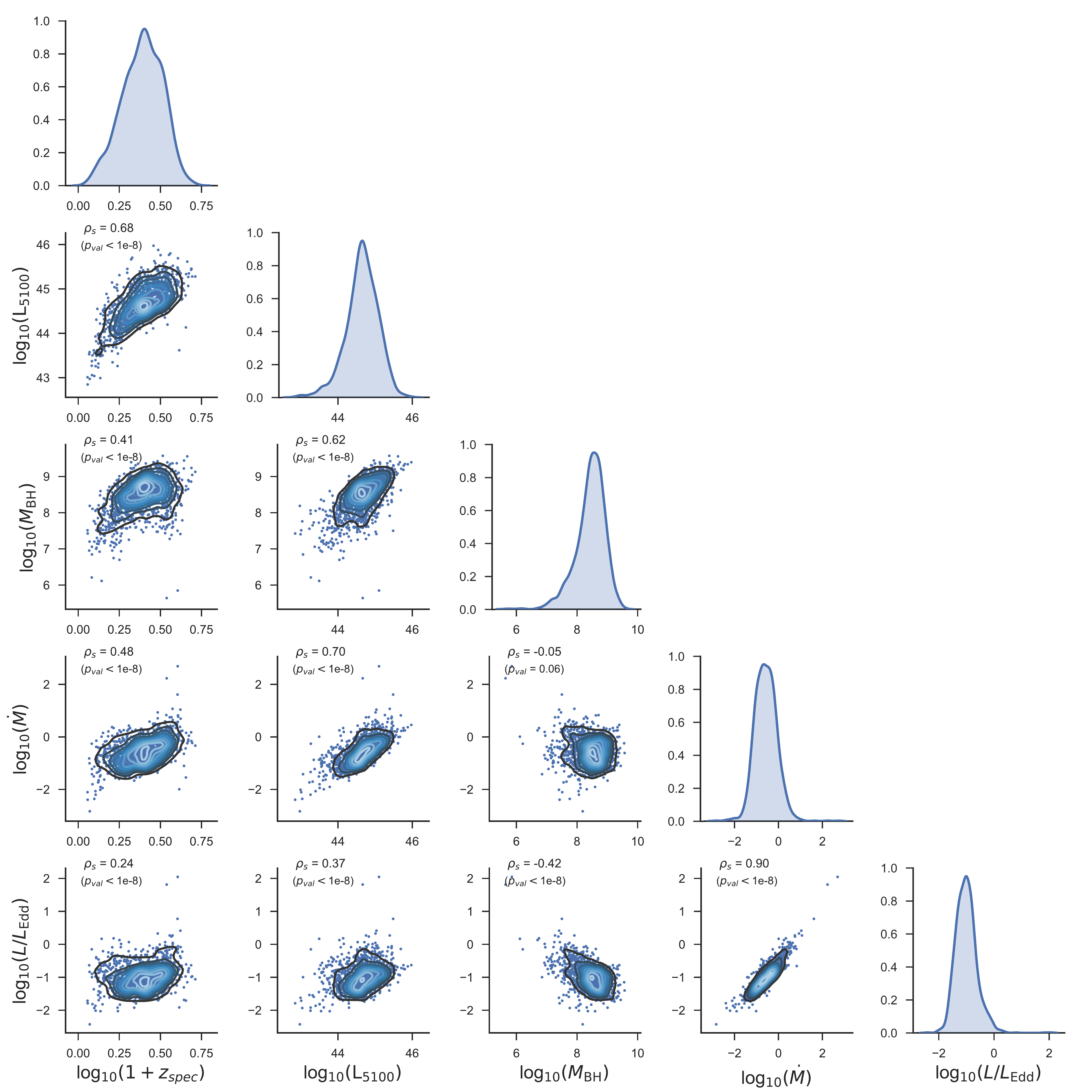}
\caption{Correlations of the spectroscopic parameters for all the sources with $M_{\text{BH}}$ and $\text{L}_{5100}$ available. The diagonal shows the individual distributions. As a reference we provide the Spearman's rank correlation coefficient for every pair of variables. \label{figure:spec_prop}}
\end{center}
\end{figure*}

Following the previous procedure, we computed $M_{\text{BH}}$ for 1899 sources, and $\text{L}_{5100}$ for 1951 sources. Figure \ref{figure:spec_prop} shows the correlations and individual distributions of $M_{\text{BH}}$, $\text{L}_{5100}$, $\dot{M}$, and $L/L_{\text{Edd}}$, for all the sources with both $M_{\text{BH}}$ and $\text{L}_{5100}$ available (1899 sources). We measure the $M_{\text{BH}}$ in units of solar masses [$M_{\odot}$], $\text{L}_{5100}$ in units of [erg s$^{-1}$], $\dot{M}$ in units of [$M_{\odot}/\text{year}$], while  $L/L_{\text{Edd}}$ is dimensionless. Our range covered for $M_{\text{BH}}$,  $\text{L}_{5100}$, and $L/L_{\text{Edd}}$ is similar than in previous variability analysis (e.g \citealt{Wilhite08,Kelly09,MacLeod10,Simm16,Caplar17}).

The strongest correlations in Figure \ref{figure:spec_prop} are: a) $\dot{M}$ and $L/L_{\text{Edd}}$, which is explained by the not particularly broad distribution of $M_{\text{BH}}$; b) $\text{L}_{5100}$ and both $M_{\text{BH}}$ an $\dot{M}$, which is related with the use of $\text{L}_{5100}$ in the determination of both quantities; and c) $z_{spec}$ with $\text{L}_{5100}$, $M_{\text{BH}}$, and $M_{\odot}$, which are mostly caused by a selection effect coming from the flux limited nature of the observations.

\begin{figure}
\begin{center}
\includegraphics[scale=0.6]{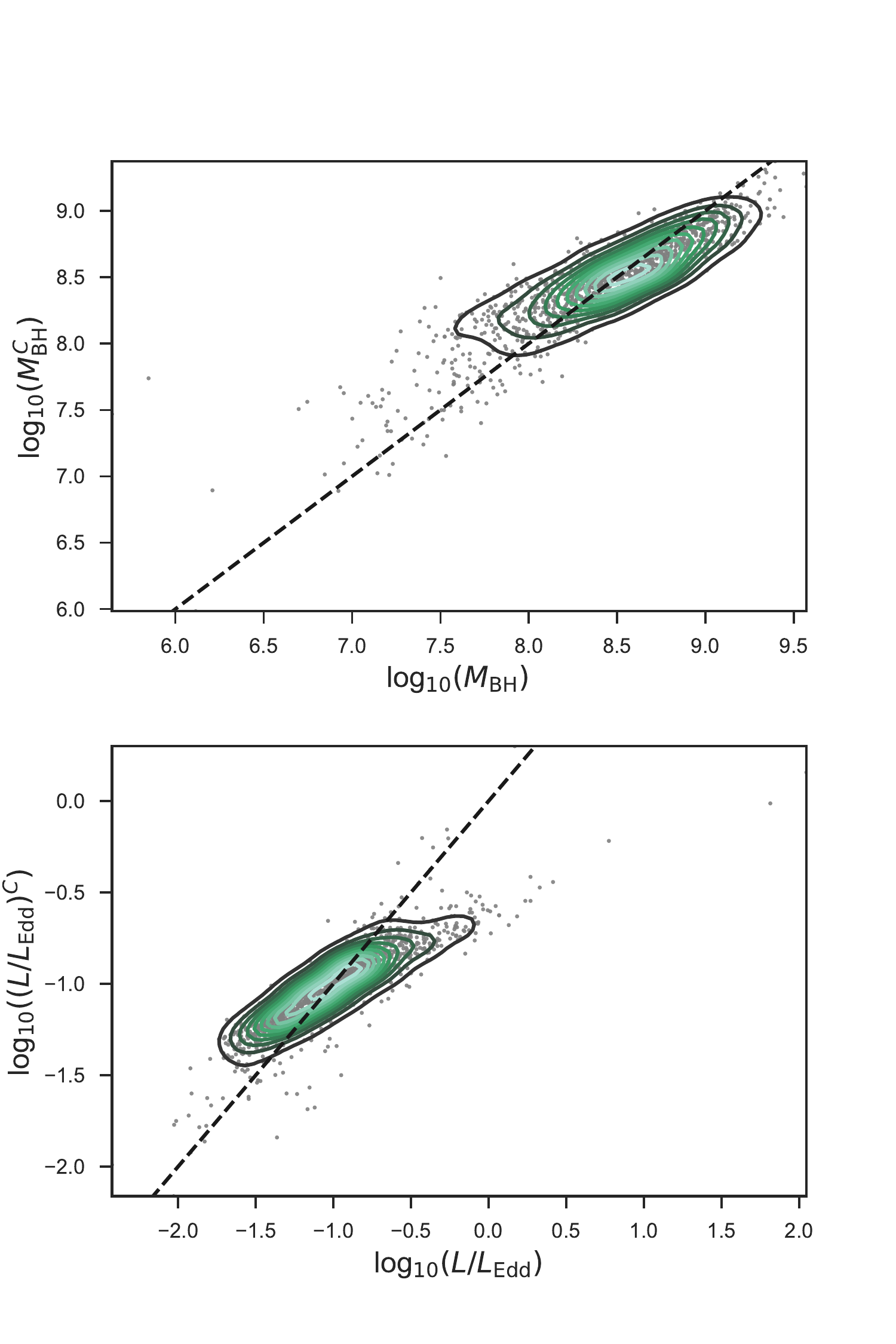}
\caption{Comparison of $M_{\text{BH}}$ vs. $M_{\text{BH}}^C$, and $L/L_{\text{Edd}}$ vs. $(L/L_{\text{Edd}})^C$. The black dashed lines shows the 1:1 relations. \label{figure:comp_corr_BH}}
\end{center}
\end{figure}

Figure \ref{figure:comp_corr_BH} shows the comparison of the $M_{\text{BH}}$ and $L/L_{\text{Edd}}$ measured using the standard single-epoch method, versus the measurements obtained using the new method proposed by \cite{MejiaRestrepo18a}. 

\section{Final sample definition}\label{sample}

Our final sample is composed by all those sources for which we could measure $M_{\text{BH}}$, $\text{L}_{5100}$, $\dot{M}$, and $L/L_{\text{Edd}}$, and have light curves with $t_{\text{rest}}\geqslant 200$ days and  $\#\text{epochs} \geqslant 40$. We have 1473 sources in our original sample that satisfy all these conditions. We call this sub-sample the ``QUEST-SDSS sample". 1348 of these sources are variable ($91.5\%$). 

We also define another sub-sample composed by all the sources with $M_{\text{BH}}$ determined from $\text{H}_{\alpha}$, $\text{H}_{\beta}$ , or Mg II line fitting. We call these sub-sample as the ``not -- C IV sample". There are 1204 sources in this sub-sample, and 1112 are variable ($92.4\%$).

The sub-sample composed by all the sources with $M_{\text{BH}}$ determined from Mg II line fitting is called the ``Mg II sample". There are 1108 sources in this sub-sample, and 1029 are variable ($92.9\%$). For 107 of the sources of this sample, the $\text{H}_{\beta}$ line was available, and used along with the Mg II line to estimate $M_{\text{BH}}$.

Finally we define the sub-sample composed by all those sources whose only available emission line is C IV. We call this sample the ``C IV sample". 236 of the 269 sources in this sub-sample are variable ($87.7\%$).

\section{Variability parameters versus physical properties}\label{var_vs_spec}

In this section we discuss the different correlations between the physical parameters measured from the SDSS spectra and the variability features measured from the QUEST light curves. For this analysis, we considered the different sub-samples defined in section \ref{sample}, but in general, we worked with the most statistically significant results, which are found for the ``not -- C IV sample".

\subsection{Bivariate correlations}\label{biv_corr}

We first analysed the bivariate correlations between the variability features and the spectral properties, using the Spearman's rank correlation coefficient ($\rho_s$), which does not consider errors measurements in the variables. It is important to remember that variability features can depend on more than one spectral property, and they would define an hyperplane which is not seen ``on-edge'' but instead through a projection onto specific axis. Therefore, some correlations can present large dispersions, even when there is a dependency of the variability feature on the spectral property.

\begin{figure*}
\begin{center}
\includegraphics[scale=0.9]{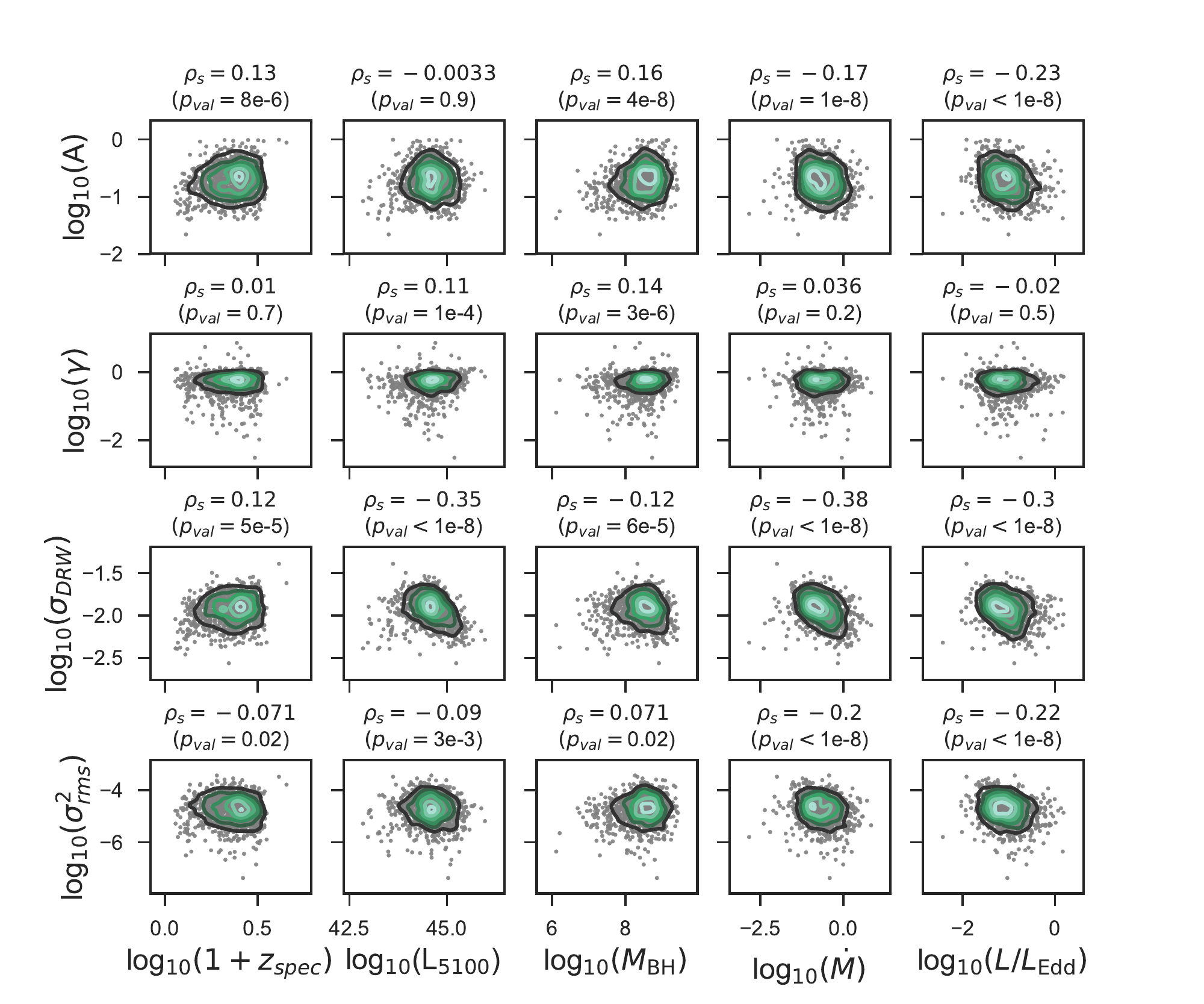}
\caption{Bivariate correlations between the variability features and the spectral properties, for the ``not -- C IV sample". The Spearman's rank correlation coefficient is shown as reference for every pair of variables. \label{figure:spec_vs_var}}
\end{center}
\end{figure*}

Figure \ref{figure:spec_vs_var} shows the bivariate correlations between the variability features and the spectral properties. From the figure we can see that $\gamma$ shows no correlation with $z_{spec}$, $\dot{M}$ and $L/L_{\text{Edd}}$, and a very weak correlation with $M_{\text{BH}}$ and $\text{L}_{5100}$. We also see that $\sigma_{DRW}$ correlates weakly with $z_{spec}$ and anti-correlates weakly with $M_{\text{BH}}$, $\text{L}_{5100}$, $\dot{M}$ and $L/L_{\text{Edd}}$. This is consistent with the findings of \cite{Kelly09}, however we have to consider the strong effect of the light curve properties in this parameter when we interpret these results. Finally, from the figure we see that  $\sigma^2_{rms}$  anti-correlates weakly with $\dot{M}$ and $L/L_{\text{Edd}}$. 

Crucially, we see that $A$ correlates weakly with $z_{spec}$ and $M_{\text{BH}}$, and anti-correlates weakly with $\dot{M}$ and $L/L_{\text{Edd}}$. These weak correlations can be driven by the large dispersion produced by correlations with other variables. Moreover, we see a lack of correlation with $\text{L}_{5100}$, which is contrary to previous findings. It must be considered that the $A$ parameter is measured for sources located at different redshifts and therefore the wavelength of rest frame emission ($\lambda_{rest}$) is different for every source. It is well known that the amplitude of the variability anti-correlates with rest frame wavelength (see S17 and references therein), which implies a positive correlation with redshift. Since $\text{L}_{5100}$ correlates with $z_{spec}$ (Figure \ref{figure:spec_prop}), the anti-correlation between $A$ and $\text{L}_{5100}$ can be hidden by the positive correlation of $A$ with redshift. Therefore, in order to detect correlations between $A$ and any physical property, instead of looking for bivariate correlations, we must perform a multivariate analysis.

\subsection{Principal Component Analysis}\label{pca}

Principal Component Analysis (PCA, \citealt{Francis99}) is a mathematical tool used to reduce the dimensionality of a data set, and it is useful to understand the correlations present in multivariate data. PCA decompose the sample into a set of linearly independent Eigenvectors that are linear combinations of the original variables. We performed a PCA on our data set, for the case of the ``not -- C IV sample", in order to see the dependencies between the different SMBH physical properties and the AGN variability features. We homogenized the data set by subtracting the mean values and normalizing by the variance. In the analysis we did not include the accretion rate ($\dot{M}$) and the excess variance ($\sigma_{rms}$), since these variables are highly correlated with the Eddington ratio ($L/L_{\text{Edd}}$) and the amplitude of the SF ($A$) respectively, and including them in the analysis produces principal components dominated by these correlations. 

Table \ref{tab:pca1} shows the results of performing a PCA on the normalized variables. We show the first five principal components (PCs). The first row gives the variances (eigenvalues) associated with every PC. The second row gives the percentage of contribution of every PC to the total variance. The third row shows the cumulative percentage of variance carried by each eigenvector. It can be seen that the first four PC together contribute $\sim 95\%$ of the variance. In Table \ref{tab:pca1} we also show for each PC the weights associated to every input variable (eigenvectors). It can be seen that the first PC is dominated by the positive correlations between redshift, luminosity and BH mass. These correlations with redshift are produced by a selection effect, since at higher redshifts our sample will naturally contain more luminous and massive sources. The second PC is dominated by the anti-correlation between $L/L_{\text{Edd}}$ and the amplitude of the variability, either measured from the SF or the DRW process. The third PC is dominated by $\gamma$, and the fourth by $L/L_{\text{Edd}}$. The fifth component is not very informative, since it contributes a small fraction of the total variance.

\begin{table}
\caption{PCA results: Eigenvectors and Eigenvalues.} \label{tab:pca1} 
\begin{center}

\begin{tabular}{lccccccc} \hline

\hline

 & PC1  & PC2 & PC3 & PC4 & PC5 \\ \hline
 
 Eigenvalue & 2.457 &  1.808 & 1.346 & 1.029 & 0.257 \\
 Percentage & 35.1\% & 25.8\% & 19.2\% &  14.7\% &  3.7\% \\
 Cumulative & 35.1\% & 60.9\% & 80.1\% & 94.8\% & 98.5 \% \\

\hline

\hline

Variable & PC1  & PC2 & PC3 & PC4 & PC5 \\ \hline

log$_{10}$(1+$z_{spec}$) &  -0.520 &  -0.118 &   0.231 &  -0.318 &  0.706 \\
log$_{10}(\text{L}_{5100}/10^{44})$ & -0.520 &  -0.387 &  -0.046 &  -0.118 &  -0.414 \\
log$_{10}(M_{\text{BH}}/10^{8})$ & -0.572 & -0.001 &   0.171 &  0.366 & -0.262 \\
log$_{10}(L/L_{\text{Edd}})$ & 0.131 &  -0.478 & -0.289 & -0.644 & -0.158 \\
log$_{10}(A)$ & -0.271 &   0.540 & -0.331 & -0.280 & -0.332 \\
log$_{10}(\gamma)$ & -0.202 &  0.213 &  -0.749 &  0.097 &  0.343 \\
log$_{10}(\sigma_{DRW})$ & -0.022 &  0.521 &  0.403 & -0.499 & -0.089 \\
 
 \hline
 
\end{tabular}
\end{center}

\end{table}

In order to have a better idea of the degree of correlation between the input variables, we computed the Spearman’s rank coefficients between the input variables and the first four PCs. The results are shown in Table \ref{tab:pca2}. We can see again that the first PC is dominated by the positive correlation between $z_{spec}$, $L_{5100}$ and $M_{\text{BH}}$. The second PC is dominated by the anti-correlation of the amplitude of the variability ($A$ and $\sigma_{DRW}$) with $L/L_{\text{Edd}}$ and $L_{5100}$. The third PC is dominated by $\gamma$, and demonstrates the positive correlation between $\gamma$ and $A$. Besides, the third PC shows a possible correlation between $\gamma$ and $L/L_{\text{Edd}}$. Finally, the fourth PC is dominated by $L/L_{\text{Edd}}$, and demonstrates an anti-correlation between $L/L_{\text{Edd}}$ and $M_{\text{BH}}$, which is expected from the definition of $L/L_{\text{Edd}}$. From these results we can conclude that the most important correlation between variability features and physical properties is for the case of the amplitude of the variability with $L/L_{\text{Edd}}$ and $L_{5100}$.

\begin{table*}
\caption{PCA results: Spearman correlation coefficients between the input variables and the four first principal components (PC1, PC2, PC3, and PC4). the p-values of the coefficients are given in parentheses.} \label{tab:pca2} 
\begin{center}

\begin{tabular}{lccccccc} \hline

\hline

Variable & PC1  & PC2 & PC3 & PC4  \\ \hline
 
 log$_{10}$(1+$z_{spec}$) &  -0.670 ($<$1e-8) &  -0.251 ($<$1e-8) &   0.161 (1e-7) &  0.006 (0.836)  \\
log$_{10}(\text{L}_{5100}/10^{44})$ & -0.808 ($<$1e-8) & -0.615 ($<$1e-8) &  -0.085 (0.005) &  0.034 (0.252)  \\
log$_{10}(M_{\text{BH}}/10^{8})$ & -0.932 ($<$1e-8) & -0.066 (0.027) &   0.243 ($<$1e-8) &  0.643 ($<$1e-8)  \\
log$_{10}(L/L_{\text{Edd}})$ & 0.192 ($<$1e-8) &  -0.712 ($<$1e-8) & -0.465 ($<$1e-8) & -0.831 ($<$1e-8)\\
log$_{10}(A)$ & -0.325 ($<$1e-8) &   0.626 ($<$1e-8) &-0.426 ($<$1e-8) & -0.045 (0.131)  \\
log$_{10}(\gamma)$ & -0.315 ($<$1e-8) &  0.275 ($<$1e-8) &  -0.802 ($<$1e-8) &  0.057 (0.058)  \\
log$_{10}(\sigma_{DRW})$ &0.130 (1e-5) &  0.591 ($<$1e-8) &  0.317 ($<$1e-8) & -0.173 ($<$1e-8)  \\

\hline
 
\end{tabular}
\end{center}

\end{table*}

\subsection{Multiple Linear Regression}\label{lin_reg}

In the previous section we showed that the amplitude of the variability anti-correlates with $L/L_{\text{Edd}}$ and $L_{5100}$, however from our PCA we cannot say whether the amplitude of the variability is mainly driven by $L/L_{\text{Edd}}$ or $L_{5100}$. Besides, is still not clear whether $\gamma$ correlates with any physical property, but from the PCA there is a possible positive correlation between $\gamma$ and $L/L_{\text{Edd}}$.

In order to have a better idea of the correlations between variability parameters and physical properties we computed Bayesian multiple linear regression. We used the Bayesian linear regression procedure of \cite{Kelly07}, which takes into account the measurement uncertainties of every variable and includes the intrinsic scatter inherent to the relation. The following sections give the results of this analysis.

\subsubsection{Trends of the amplitude of the SF with physical properties}\label{lin_reg_A}

In the previous section we showed that $A$ presents an anti-correlation with $L/L_{\text{Edd}}$ and $L_{5100}$, a multiple linear regression analysis can help us to differentiate which physical property drives these anti-correlations. 

S17 showed that there is a positive correlation between $A$ with $z_{spec}$, which is produced by an anti-correlation between $A$ with $\lambda_{rest}$ (see Figure 11 in S17). Therefore, given the wide range in redshift of our sample, we must always consider the correlation with redshift when we analyse correlations with any other physical parameter.

Table \ref{tab:reg_A_noCIV} shows the results of the Bayesian multiple linear regression for $A$ as the dependent variable, and different combinations of the spectral properties as independent variables. In the table, every column gives the value of the intercept ($\alpha$), the slope ($\beta$) associated with a given physical property, and the intrinsic scatter associated to the regression model ($\epsilon$). When the value of the slope is replaced by X, it means that the parameter was not included in the regression model. 

\begin{table*}
\caption{Linear Regression $\alpha$, $\beta$ and $\epsilon$ coefficients for $A$ as the dependent variable (for the not -- C IV sample). The columns headed by physical quantities refer to their slope in the regression model ($\beta$).} \label{tab:reg_A_noCIV} 
\begin{center}

\begin{tabular}{c|ccccccc} \hline

$\#$ & $\alpha$& log$_{10}$(1+$z_{spec}$) & log$_{10}(\text{L}_{5100}/10^{44})$ & log$_{10}(M_{\text{BH}}/10^{8})$& log$_{10}(L/L_{\text{Edd}})$ & log$_{10}(\dot{M})$ & $\epsilon$\\ \hline

\toprule

 1 &  -0.88 $\pm$ 0.03 &0.44 $\pm$ 0.08 & X & X & X & X & 0.26 $\pm$ 0.01\\  
 
  2 &   -0.74 $\pm$ 0.01 & X & 0.03 $\pm$ 0.02  & X & X & X & 0.26 $\pm$ 0.01 \\
  
  3 &  -0.77 $\pm$ 0.01 & X & X &  0.12 $\pm$ 0.02 & X  & X &    0.25 $\pm$ 0.01 \\
  
 4 & -0.93 $\pm$ 0.03 & X & X & X &   -0.19 $\pm$ 0.02 & X &  0.25 $\pm$ 0.01 \\
  
 5 &  -0.79 $\pm$ 0.01 &  X & X & X & X & -0.09 $\pm$ 0.02 & 0.26 $\pm$ 0.01\\ 
 
   \hline
    
   6 &    -0.87 $\pm$ 0.03  &   0.52 $\pm$ 0.11 & -0.21 $\pm$ 0.03 & 0.18 $\pm$ 0.03 & X & X & 0.25 $\pm$ 0.01  \\
     
   7 &-1.08 $\pm$ 0.04 & 0.50 $\pm$ 0.11  &  -0.02 $\pm$ 0.03 & X  &  -0.18 $\pm$ 0.03 & X  & 0.25 $\pm$ 0.01      \\ 
    
   8 &   -1.14 $\pm$ 0.06  &     0.57 $\pm$ 0.11  & X &  -0.04 $\pm$ 0.03 & -0.22 $\pm$ 0.03 & X  & 0.25 $\pm$ 0.01      \\

\hline
    
   9 &  -0.92 $\pm$ 0.03 &  0.72 $\pm$ 0.10 &  -0.10 $\pm$ 0.03 & X & X & X & 0.25 $\pm$ 0.01 \\
    
   10 &  -0.83 $\pm$ 0.03 & 0.18 $\pm$ 0.10 & X &   0.09 $\pm$ 0.02 & X & X & 0.25 $\pm$ 0.01 \\
    
   \textbf{11} & \textbf{-1.09 $\pm$ 0.04} &  \textbf{ 0.45 $\pm$ 0.07} & \textbf{X} &\textbf{ X} & \textbf{-0.19 $\pm$ 0.02} & \textbf{X} &  \textbf{ 0.25 $\pm$ 0.01 }\\

\hline
\end{tabular}
\end{center}

\end{table*}

Regressions \#1 to \#5 in Table \ref{tab:reg_A_noCIV} correspond to models with one single independent variable. We see that the most significant correlations are for $z_{spec}$, $L/L_{\text{Edd}}$ and $M_{\text{BH}}$. Since $\dot{M}$ and $L/L_{\text{Edd}}$ are highly correlated (see Figure \ref{figure:spec_prop}), and including these two variables together can produce multicollinearity in the regression model, we decided to exclude $\dot{M}$ from the regression models, and keep $L/L_{\text{Edd}}$. 

For the regressions \#6 to \#10, we decided to include always $z_{spec}$ as one of the independent variables, because we are analysing light curves observed in a fixed photometric band (Q), which implies that the rest frame wavelength of every light curve will depend on the redshift of the source. Besides, we do not include in Table \ref{tab:reg_A_noCIV} a regression model with $z_{spec}$, $\text{L}_{5100}$, $M_{\text{BH}}$, and $L/L_{\text{Edd}}$ as independent variables, because the multicollinearity of the variables does not allow the Bayesian method to converge and return confident regression coefficients.

Regression \#6 shows that when the model includes $z_{spec}$, $M_{\text{BH}}$ and $\text{L}_{5100}$, the slopes for $M_{\text{BH}}$ and $\text{L}_{5100}$ satisfy (within 1 $\sigma$) the relation: $\beta_{\text{L}_{5100}} \sim -\beta_{M_{\text{BH}}}$. This would be expected if $L/L_{\text{Edd}}$ is the driver of the amplitude variability.

Regression \#11 corresponds to a model which includes both the Eddington ratio and redshift. We can see that the coefficients are statistically significant, and thus we propose this model as the best regression model for the amplitude of the variability. This can be confirmed when we see regressions \#7 and \#8,  where adding $\text{L}_{5100}$ or $M_{\text{BH}}$ in the model, besides $L/L_{\text{Edd}}$ and redshift, gives statistically insignificant slopes for $M_{\text{BH}}$ or $\text{L}_{5100}$.

\begin{figure*}
\begin{center}
\includegraphics[scale=0.5]{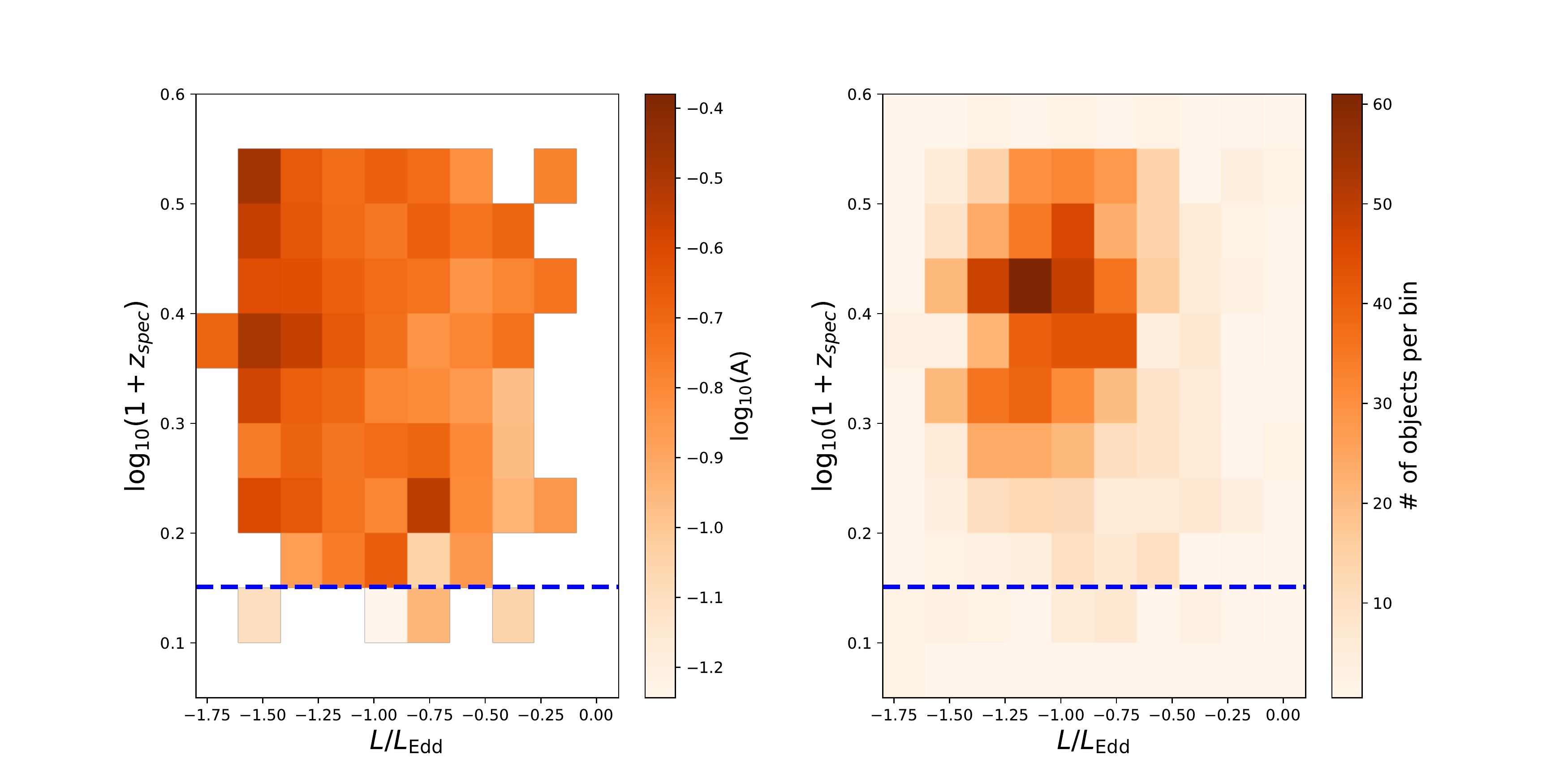}
\caption{Left: Mean value of $A$ in a grid of $z_{spec}$ and $L/L_{\text{Edd}}$, for bins with 3 or more sources.  Right: number of sources per bin of $z_{spec}$ and $L/L_{\text{Edd}}$. The red dashed line shows the redshift from which we have available the Mg II line in the SDSS spectra (0.42). \label{figure:grid_plot}}
\end{center}
\end{figure*}

Figure \ref{figure:grid_plot} shows the dependency of $A$ with  $z_{spec}$ and $L/L_{\text{Edd}}$. We can see the anti-correlation of $A$ with $L/L_{\text{Edd}}$, and the positive correlation of $A$ with redshift. The trend with $L/L_{\text{Edd}}$ is more significant.

In order to break the effects of the $A-\gamma$ degeneracy, we re-computed the regression model \#11 of Table \ref{tab:reg_A_noCIV}, but considering only variable sources whose measured $\gamma$ range between $\gamma_{med} \pm 0.1$, where $\gamma_{med}$ corresponds to the median value of $\gamma$ measured for the well-sampled sub-sample (0.53), i.e. we consider those sources with $0.43 \leq \gamma \leq 0.63$. Selecting a narrow range in measured $\gamma$ allows us to better discriminate between different \emph{intrinsic} values of the amplitude A, as can be seen in Figure \ref{figure:sim_A_vs_g}. There are 322 variables sources from the not -- C IV sample in this range of $\gamma$. In this case, the results of the regression are: $\alpha=-1.06 \pm 0.05 $, $\beta_{\text{log}_{10}(1+z_{spec})}=0.45\pm0.1$, $\beta_{\text{log}_{10}(L/L_{\text{Edd}})}=-0.22\pm 0.03$, and $\epsilon=0.14\pm 0.01$. The results for the slopes of $z_{spec}$ and $L/L_{\text{Edd}}$ are consistent with what we found for the whole not -- C IV sample, at $1\sigma$ level. However, there is a considerable reduction in the intrinsic scatter measured for the reduced sample. This implies that a large part of the scatter measured in the different regression models of Table \ref{tab:reg_A_noCIV} comes from the $A-\gamma$ degeneracy.

\begin{table*}
\caption{Linear Regression $\alpha$, $\beta$ and $\epsilon$ coefficients for $A$ as the dependent variable (for standard black hole masses). The columns headed by physical quantities refer to their slope in the regression model ($\beta$).} \label{tab:regA_standardBHmass} 
\begin{center}

\begin{tabular}{c|ccccccc} \hline

sample & $\alpha$ & log$_{10}$(1+$z_{spec}$) & log$_{10}(L/L_{\text{Edd}})$ & $\epsilon$\\ 

\hline
QUEST-SDSS &  -1.01 $\pm$ 0.03 &   0.41 $\pm$ 0.06 & -0.13 $\pm$ 0.02 & 0.25 $\pm$ 0.01 \\
not -- C IV  & -1.09 $\pm$ 0.04 &    0.45 $\pm$ 0.07 &  -0.19 $\pm$ 0.02 &  0.21 $\pm$ 0.01  \\
Mg II & -1.02 $\pm$ 0.04 &   0.23 $\pm$ 0.09 &     -0.21 $\pm$ 0.03 &  0.24 $\pm$ 0.01 \\
C IV &  -0.47 $\pm$ 0.20 &     -0.37 $\pm$ 0.36 & 0.02 $\pm$ 0.04 &   0.25 $\pm$ 0.01 \\

\hline

sample & $\alpha$ & log$_{10}$(1+$z_{spec}$) &  log$_{10}(\text{L}_{5100}/10^{44})$ &log$_{10}(M_{\text{BH}}/10^8)$ & $\epsilon$\\ 

\hline
QUEST-SDSS &   -0.86  $\pm$ 0.02 &  0.52  $\pm$ 0.08 &  -0.17  $\pm$ 0.03 & 0.13  $\pm$ 0.02 &  0.25  $\pm$ 0.01 \\
not -- C IV &  -0.87 $\pm$ 0.03  &  0.52 $\pm$ 0.11  &  -0.21 $\pm$ 0.03 & 0.18 $\pm$ 0.03 &  0.25 $\pm$ 0.01 \\
Mg II &  -0.79  $\pm$ 0.04 &  0.43  $\pm$ 0.11 &   -0.26  $\pm$ 0.03 &    0.18  $\pm$ 0.03 &  0.24  $\pm$ 0.01  \\
C IV &  -0.51  $\pm$ 0.20 &  -0.16  $\pm$ 0.34 &    -0.08  $\pm$ 0.09 &  -0.02  $\pm$ 0.04 &  0.25  $\pm$ 0.01 \\

\hline
\end{tabular}
\end{center}
\end{table*}

Table \ref{tab:regA_standardBHmass} gives the linear regression coefficients for $z_{spec}$ and $L/L_{\text{Edd}}$ considering the different samples defined in section \ref{sample}. From the table we can see that the results for the C IV sample are different from the results obtained for the other three samples. For the QUEST-SDSS sample, we can see a decrement in the slope of $L/L_{\text{Edd}}$, produced by the presence of sources with C IV measurements. The slope of $L/L_{\text{Edd}}$ is consistent for the not -- C IV and Mg II samples, but the slope for $z_{spec}$ changes. This can be produced by the reduced dynamic range in redshift for the Mg II sample, in comparison with the not -- C IV sample. Moreover, from Figure \ref{figure:grid_plot} we can see that when we only consider sources from the Mg II sample (sources above the red dashed line), we lose most of the sources with low variability.

Table \ref{tab:regA_standardBHmass} also shows the linear regression coefficient for a model with redshift, $\text{L}_{5100}$ and $M_{\text{BH}}$. We can see that, as we showed in Table \ref{tab:reg_A_noCIV}, for the not -- C IV sample the slopes of $\text{L}_{5100}$ and $M_{\text{BH}}$ satisfy $\beta_{\text{L}_{5100}} \sim - \beta_{M_{\text{BH}}}$. However, for the QUEST-SDSS and Mg II samples, the relation between the slopes is not so evident. Again, for the case of the QUEST-SDSS sample we have contamination from C IV. The difference between the results for the Mg II and not -- C IV samples can be driven by the change in the dynamic range of $z_{spec}$ (see Figure \ref{figure:spec_prop}), and the strong correlation between $z_{spec}$ and $\text{L}_{5100}$ (produced by a selection effect), since the slope for $z_{spec}$ also increases when we include $\text{L}_{5100}$ in the model.

In order to see whether smaller ranges of redshift can reduce the effects in the regression analysis of the $z_{spec}$ versus $L_{5100}$ correlation, we computed the regression models again, but considering sources from the not -- C IV sample with $1.5 \leq z_{spec} \leq 1.8$, since in this range of redshift the correlation between $z_{spec}$ and $L_{5100}$ is smaller ($\rho_s=0.12$, $p_{val}=0.09$). We found similar results, but the results are less statistically significant due to the low number of sources considered (there are 213 variable sources in this range of redshift). For the case of the regression model with $z_{spec}$, $L_{5100}$ and $M_{\text{BH}}$, the results of the regression are: $\alpha=0.03 \pm 0.55 $, $\beta_{\text{log}_{10}(1+z_{spec})}=1.51\pm1.30$, $\beta_{\text{log}_{10}(L_{5100})}=-0.20\pm 0.08$,  $\beta_{\text{log}_{10}(M_{\text{BH}})}=-0.12\pm 0.07$, and $\epsilon=0.21\pm 0.01$. And for the case of the regression with $z_{spec}$ and $L/L_{\text{Edd}}$, the results of the regression are: $\alpha=-0.22 \pm 0.56 $, $\beta_{\text{log}_{10}(1+z_{spec})}=1.50\pm1.29$, $\beta_{\text{log}_{10}(L/L_{\text{Edd}})}=-0.15\pm 0.06$, and $\epsilon=0.21\pm 0.01$. From these results, we can see that $\beta_{\text{log}_{10}(L_{5100})} \sim - \beta_{\text{log}_{10}(M_{\text{BH}})}$ at $1\sigma$ level, and that the anti-correlation between $A$ and $(L/L_{\text{Edd}})$ is still present.

\begin{table*}
\caption{Linear Regression $\alpha$, $\beta$ and $\epsilon$ coefficients for $A$ as the dependent variable (for corrected black hole masses). The columns headed by physical quantities refer to their slope in the regression model ($\beta$).} \label{tab:regA_correctedBHmass} 
\begin{center}

\begin{tabular}{c|ccccccc} \hline

sample & $\alpha$ & log$_{10}$(1+$z_{spec}$) & log$_{10}((L/L_{\text{Edd}})^C)$  & $\epsilon$\\ 

\hline
QUEST-SDSS &   -1.19 $\pm$ 0.08 &    0.48 $\pm$ 0.07 &  -0.28 $\pm$ 0.06 & 0.25 $\pm$ 0.01 \\
not -- C IV  &  -1.22 $\pm$ 0.08 &    0.56 $\pm$ 0.08 &    -0.28 $\pm$ 0.06&    0.25 $\pm$ 0.01 \\
Mg II &  -1.16 $\pm$ 0.12 & 0.36 $\pm$ 0.10 &  -0.30 $\pm$ 0.08 &   0.25 $\pm$ 0.01 \\
C IV &  4.33 $\pm$ 20.35 &  -2.08 $\pm$ 11.80 &   4.11 $\pm$ 21.21 & 0.23 $\pm$ 0.04 \\

\hline

sample & $\alpha$ & log$_{10}$(1+$z_{spec}$) &  log$_{10}(\text{L}_{5100}/10^{44})$ &log$_{10}(M_{\text{BH}}^C/10^8)$  & $\epsilon$\\ 

\hline
QUEST-SDSS &     -0.84 $\pm$ 0.03 &   0.42 $\pm$ 0.09 &   -0.29 $\pm$ 0.05 &   0.31 $\pm$ 0.06 &0.25 $\pm$ 0.01 \\
not -- C IV &  -0.88 $\pm$ 0.03 &   0.55 $\pm$ 0.11 &  -0.28 $\pm$ 0.05 & 0.29 $\pm$ 0.06 &   0.25 $\pm$ 0.01 \\
Mg II & -0.81 $\pm$ 0.04  &  0.48 $\pm$ 0.12 &   -0.27 $\pm$ 0.06 &  0.20 $\pm$ 0.08 & 0.25 $\pm$ 0.01 \\
C IV &  -0.59 $\pm$ 2.2 &   -0.33 $\pm$ 4.23 & 2.00 $\pm$ 6.28 &   -2.73 $\pm$ 8.26 &  0.23 $\pm$ 0.04\\

\hline
\end{tabular}
\end{center}
\end{table*}

As mentioned in section \ref{spec_fit}, \cite{MejiaRestrepo18a} proposed new corrections for the estimation of $M_{\text{BH}}$. We show in Table \ref{tab:regA_correctedBHmass} the linear regression coefficients for a model with $z_{spec}$ and $(L/L_{\text{Edd}})^C$, and for a model with $z_{spec}$, $\text{L}_{5100}$, and $M_{\text{BH}}^C$, for the different samples of section \ref{sample}. We can see that the slopes for $z_{spec}$ and $(L/L_{\text{Edd}})^C$ increase for the QUEST-SDSS, not -- C IV, and C IV samples, but the errors in the slopes also increase. The slopes for  $(L/L_{\text{Edd}})^C$ are more similar for the different samples than the slopes for $z_{spec}$. This can be related with the difference in the dynamic ranges of $z_{spec}$ for different samples. For the regression model with $z_{spec}$, $\text{L}_{5100}$, and $M_{\text{BH}}^C$, the slopes of $\text{L}_{5100}$ and $M_{\text{BH}}^C$ satisfy the relation $\beta_{\text{L}_{5100}} \sim - \beta_{M_{\text{BH}}^C}$ for the QUEST-SDSS, not -- C IV and Mg II samples (at $1\sigma$). This supports our idea that $L/L_{\text{Edd}}$ is the driver of the variability amplitude. The difference of these results with what we showed in Table \ref{tab:regA_standardBHmass} can be given by the reduction of the scatter in the determined black hole mass when we use the corrections proposed by \cite{MejiaRestrepo18a}.

\begin{table}
\caption{Linear Regression $\alpha$, $\beta$ and $\epsilon$ coefficients for $A$ as the dependent variable, for spectral properties derived from Mg II. The columns headed by physical quantities refer to their slope in the regression model ($\beta$).} \label{tab:regA_MgII} 
\begin{center}

\begin{tabular}{cccc} \hline

 $\alpha$ & log$_{10}$(1+$z_{spec}$) & log$_{10}$(L$_{\text{Mg II}}$/L$_{3000}$) & $\epsilon$\\ 
  0.02 $\pm$ 0.09 &    0.29 $\pm$ 0.09 &  0.44 $\pm$ 0.05 & 0.24 $\pm$ 0.01 \\

\hline

 $\alpha$ & log$_{10}$(1+$z_{spec}$) & log$_{10}$(L$_{5100}$/L$_{3000}$) & $\epsilon$\\ 
  -0.67 $\pm$ 0.21 &  -0.55 $\pm$ 0.93  & -0.19 $\pm$ 0.13  & 0.22 $\pm$ 0.02 \\
  
  \hline
  
 $\alpha$ & log$_{10}$(1+$z_{spec}$) & log$_{10}$(L$_{3000}$/L$_{1450}$) & $\epsilon$\\ 
  -0.64 $\pm$ 0.26 &   -0.06 $\pm$ 0.55 &  0.21 $\pm$ 0.15  & 0.27 $\pm$ 0.01 \\

\hline
\end{tabular}
\end{center}
\end{table}

Table \ref{tab:regA_MgII} shows three linear regression models that consider different spectral properties of the Mg II line as the independent variable. The first regression model corresponds to a model with L$_{\text{Mg II}}$/L$_{3000}$, which is a proxy of the EW of the line. Previous analysis have found that there is a strong anti-correlation between the equivalent width of Mg II and $L/L_{\text{Edd}}$ (see \citealt{Netzer13} and references therein). Our results shows a positive correlation between L$_{\text{Mg II}}$/L$_{3000}$ and $A$, which supports our interpretation that $L/L_{\text{Edd}}$ is the driver of the amplitude.

The second and third regression models of Table \ref{tab:regA_MgII} include the spectral slopes $\text{L}_{5100}/\text{L}_{3000}$ and $\text{L}_{3000}/\text{L}_{1450}$. We can see that there is not statistically significant correlation between the amplitude of the variability and these spectral slopes.

\begin{table*}
\caption{Linear Regression $\alpha$, $\beta$ and $\epsilon$ coefficients for $A$ as the dependent variable (spectral properties per emission line). The columns headed by physical quantities refer to their slope in the regression model ($\beta$). } \label{tab:regA_FWHM} 
\begin{center}

\begin{tabular}{c|ccccccc} \hline

 line & $\alpha$ & log$_{10}$(1+$z_{spec}$) & log$_{10}$(FWHM) & log$_{10}(\lambda L_{\lambda}/10^{44}$) & $\epsilon$\\ \hline
H$_{\alpha}$ (57) &  -8.76 $\pm$ 5.20 &    1.29 $\pm$ 1.65 &   0.39 $\pm$ 0.16 &  0.14 $\pm$ 0.12 &  0.26 $\pm$ 0.03 \\

H$_{\beta}$ (172) & -2.76 $\pm$ 2.31 & 1.62 $\pm$ 0.46 & 0.23 $\pm$ 0.08 & 0.02 $\pm$ 0.05 & 0.24 $\pm$ 0.01 \\
  Mg II (1063) &  3.50 $\pm$ 1.11 &     0.44 $\pm$ 0.11 &    0.34 $\pm$ 0.05 &   -0.12 $\pm$ 0.02 &  0.24 $\pm$ 0.01 \\
  C IV (460) &    3.60 $\pm$ 1.93 & 0.02 $\pm$ 0.23 &   -0.08 $\pm$ 0.08 &-0.09 $\pm$ 0.04 & 0.26 $\pm$ 0.01 \\ 

\hline

\multicolumn{5}{l}{\textbf{Note}. In parentheses we show the number of variable sources considered per line.}
\end{tabular}
\end{center}
\end{table*}

We looked for correlations between $A$ and parameters derived from the line fitting. Table \ref{tab:regA_FWHM} shows the regression coefficients for models that consider the FWHM and continuum luminosity ($\lambda L_{\lambda}$) for the H$_{\alpha}$, H$_{\beta}$, Mg II and C IV lines. For the case of H$_{\alpha}$, the statistics is poor given the low number of variable sources with this line available in the SDSS spectra. Despite that, we can see a positive correlation between $A$ and FWHM(H$_{\alpha}$). For the case of H$_{\beta}$, the results are similar, with a positive correlation between $A$ and FWHM(H$_{\beta}$). Mg II has the best statistics, with 1063 variable sources available. In this case we also see a correlation between $A$ and FWHM(Mg II), but also an anti-correlation between $A$ and $\text{L}_{3000}$. The results for C IV are completely different, with no significant correlation between $A$ and FWHM(C IV) or $\text{L}_{1450}$. This can be related with the known problems of using the C IV line to measure black hole masses, since the line profile deviates considerably from Keplerian-type motion, and can be influenced by winds emanating from the accretion disk (see \citealt{Netzer13,MejiaRestrepo18b},  and references therein). 

The positive correlations between $A$ and the FWHM of H$_{\alpha}$, H$_{\beta}$, and Mg II are expected for a variability process whose amplitude is driven by $L/L_{\text{Edd}}$, since $L/L_{\text{Edd}} \propto \text{FWHM}^{-2} (\lambda L_{\lambda})^{1-\alpha}$ (following the equations of section \ref{spec_fit}). Under this assumption, the anti-correlation between $A$ and $\text{L}_{3000}$ is also expected. The lack of correlation between $A$ and $\text{L}_{6200}$, and $\text{L}_{5100}$ can be given by the differences in the continuum luminosity range covered by these lines compared to Mg II.

\begin{table*}
\caption{Linear Regression $\alpha$, $\beta$ and $\epsilon$ coefficients for other amplitude features as dependent variables (not -- C IV sample). The columns headed by physical quantities refer to their slope in the regression model ($\beta$).} \label{tab:reg_otherAmp} 
\begin{center}

\begin{tabular}{c|ccccccc} \hline

feature &  $\alpha$ & log$_{10}$(1+$z_{spec}$) & log$_{10} (L/L_{\text{Edd}})$ & $\epsilon$\\ 
  
\hline

$\sigma_{DRW}$ &-2.20 $\pm$ 0.02  &     0.26 $\pm$ 0.04 &    -0.16 $\pm$ 0.01 & 0.12 $\pm$ 0.004  \\

$\sigma_{rms}^2$ &  -5.07 $\pm$ 0.07 & -0.13 $\pm$ 0.14 &  -0.34 $\pm$ 0.05 &  0.48 $\pm$  0.01 \\
   
\hline

feature & $\alpha$ & log$_{10}$(1+$z_{spec}$) &  log$_{10}(\text{L}_{5100}/10^{44})$ &log$_{10}(M_{\text{BH}}/10^8)$ & $\epsilon$\\ \hline

 $\sigma_{DRW}$ & -2.13 $\pm$ 0.01 & 1.04 $\pm$ 0.05 & -0.33 $\pm$ 0.01 &    0.06 $\pm$ 0.01 & 0.09 $\pm$ 0.0034 \\
 
 $\sigma_{rms}^2$ & -4.64 $\pm$ 0.06 & -0.31 $\pm$ 0.21 &  -0.29 $\pm$ 0.06 & 0.35 $\pm$ 0.05 & 0.48 $\pm$ 0.01  \\
 
\hline

\end{tabular}
\end{center}

\end{table*}

\subsubsection{Trends of other amplitude features with physical properties}\label{lin_reg_other_amp}

We tested whether $\sigma_{rms}^2$ and $\sigma_{DRW}$ also show correlations with $z_{spec}$ and $L/L_{\text{Edd}}$ or $\text{L}_{5100}$ and $M_{\text{BH}}$. The results are shown in Table \ref{tab:reg_otherAmp}. For the case of $\sigma_{rms}^2$, we see a significant anti-correlation with $L/L_{\text{Edd}}$, which is consistent with what we found for $A$. We also found a lack of significant correlation with $z_{spec}$. This can be given by the positive correlation between the amplitude of the variability and $z_{spec}$ and the negative correlation between the length of the light curve and $z_{spec}$. Since $\sigma_{rms}^2$ considers the variance of the whole light curve, for sources at high redshift we observe shorter light curves than at low redshift, and therefore the correlation with $z_{spec}$ is considerably diminished.

For the case of $\sigma_{DRW}$, we see an anti-correlation with $L/L_{\text{Edd}}$ and a positive correlation with $z_{spec}$. This is in contrast with the results reported by \cite{Kelly09}, who found no correlation between $L/L_{\text{Edd}}$ and $\sigma_{DRW}$. This can be given by the strong dependency of $\sigma_{DRW}$ on the sampling of the light curve, and the considerably small number of sources, with respect to our sample, used by \cite{Kelly09}. Our results also show that $\sigma_{DRW}$ correlates negatively with $\text{L}_{5100}$ and has no correlation with $M_{\text{BH}}$. \cite{Kelly09} found a similar slope for $\text{L}_{5100}$ for their model with $z_{spec}$ included (see their Eq. 25). Since we found no correlation with $M_{\text{BH}}$, we propose that the anti-correlation between $\sigma_{DRW}$ and  $L/L_{\text{Edd}}$ is given by the anti-correlation between $\sigma_{DRW}$  and $\text{L}_{5100}$. We must consider the implication of these results with caution, since $\sigma_{DRW}$ is strongly affected by the light curve sampling. Particularly, Figure \ref{figure:varfeat_lcprop} shows that $\sigma_{DRW}$ anti-correlates with the number of epochs and the length of the light curve. More luminous sources have higher probabilities to be detected in more epochs than fainter sources. In fact, the Spearman's rank correlation coefficient for $\text{L}_{5100}$ and the number of epochs is 0.45 ($p_{val}=$1e-4). Therefore, the anti-correlation between $\sigma_{DRW}$  and $\text{L}_{5100}$ can be just a reflection of the anti-correlation between $\sigma_{DRW}$  and the number of epochs.

\subsubsection{Trends of the logarithmic gradient of the variability ($\gamma$) with physical properties}\label{lin_reg_g}

In section \ref{biv_corr} we showed that $\gamma$ correlates very weakly with $\text{L}_{5100}$ and $M_{\text{BH}}$ (see Figure \ref{figure:spec_vs_var}), and from the PCA there is evidence of a positive correlation between $\gamma$ and $L/L_{\text{Edd}}$. In order to test whether any of these correlations exists, we performed a linear regression analysis. Table \ref{tab:reg_gamma} shows the linear regression coefficients for $\gamma$, when we consider spectral features as single independent variables in the regression model, for the case of the not -- C IV sample. We can see that $\gamma$ does not have statistically significant correlation with any physical parameter, since the absolute values of the slopes for every the regression model are small (lower than 0.1) and/or have high errors compared to the measured values. 

Some sources have values of $\gamma$ that are inconsistent with a DRW process. In section \ref{sim_SF} we showed that if a measured value of $\gamma$ range between 0.0 and 0.75, we cannot discard a DRW process as the best model to describe the variability. In the well-sampled sub-sample, 325 of the 1579 variable sources have values of $\gamma$ higher than 0.75 ($20.6\%$ of the sample). For these sources, the value of $\gamma$ differs considerably from 0.5, and therefore, a DRW model is not sufficient to model the variability. When we compare the distributions of the SMBH physical properties of a) the 325 sources with $\gamma>0.75$, and b) the rest of the sample; we do not observe any difference between the populations. We also do not observe differences in the light curve sampling of these two populations, we can therefore discard an observational bias in the distribution of $\gamma$.

\begin{table}
\caption{Linear Regression $\alpha$, $\beta$ and $\epsilon$ coefficients for $\gamma$ as the dependent variable (not -- C IV). The columns headed by physical quantities refer to their slope in the regression model ($\beta$).} \label{tab:reg_gamma} 
\begin{center}

\begin{tabular}{cccccccc} \hline

 $\alpha$ & log$_{10}$(1+$z_{spec}$) & $\epsilon$\\ 
 -0.32 $\pm$ 0.03  &    0.01 $\pm$ 0.08 &    0.28 $\pm$ 0.01 \\

\hline

 $\alpha$ & log$_{10}(\text{L}_{5100}/10^{44})$ & $\epsilon$\\ 
-0.37 $\pm$ 0.01  &     0.09 $\pm$ 0.02 &     0.28 $\pm$ 0.01  \\
  
\hline
    
$\alpha$ & log$_{10}(M_{\text{BH}}/10^8)$ & $\epsilon$\\ 
   -0.35 $\pm$ 0.01 &     0.08 $\pm$ 0.02 &   0.28 $\pm$ 0.01  \\

\hline
    
 $\alpha$ &log$_{10} (L/L_{\text{Edd}})$ & $\epsilon$\\ 
  -0.34 $\pm$ 0.03  &    -0.01 $\pm$ 0.03 &   0.28 $\pm$ 0.01 \\

\hline
    
$\alpha$ & log$_{10}(\dot{M})$ & $\epsilon$\\ 
 -0.30 $\pm$ 0.02  &   0.03 $\pm$ 0.02 &  0.28 $\pm$ 0.01 \\

\hline
\end{tabular}
\end{center}

\end{table}

\subsection{Differences between variable and non-variable sources}

\begin{figure}
\begin{center}
\includegraphics[scale=0.47]{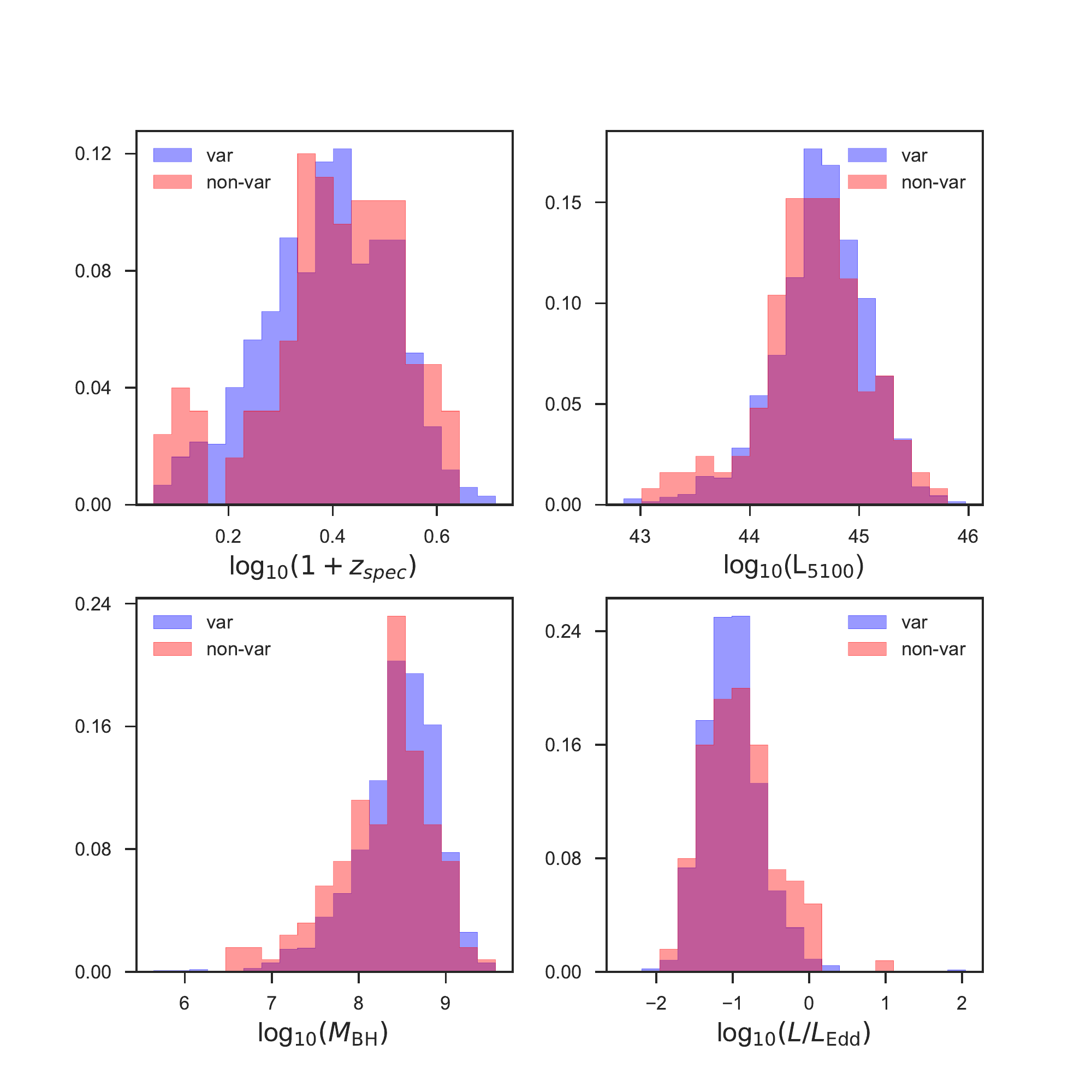}
\caption{Normalized histogram of the physical properties of variable (blue) and non-variable (red) sources from the QUEST-SDSS sample. \label{figure:hist_var_novar_spec}}
\end{center}
\end{figure}

\begin{figure}
\begin{center}
\includegraphics[scale=0.7]{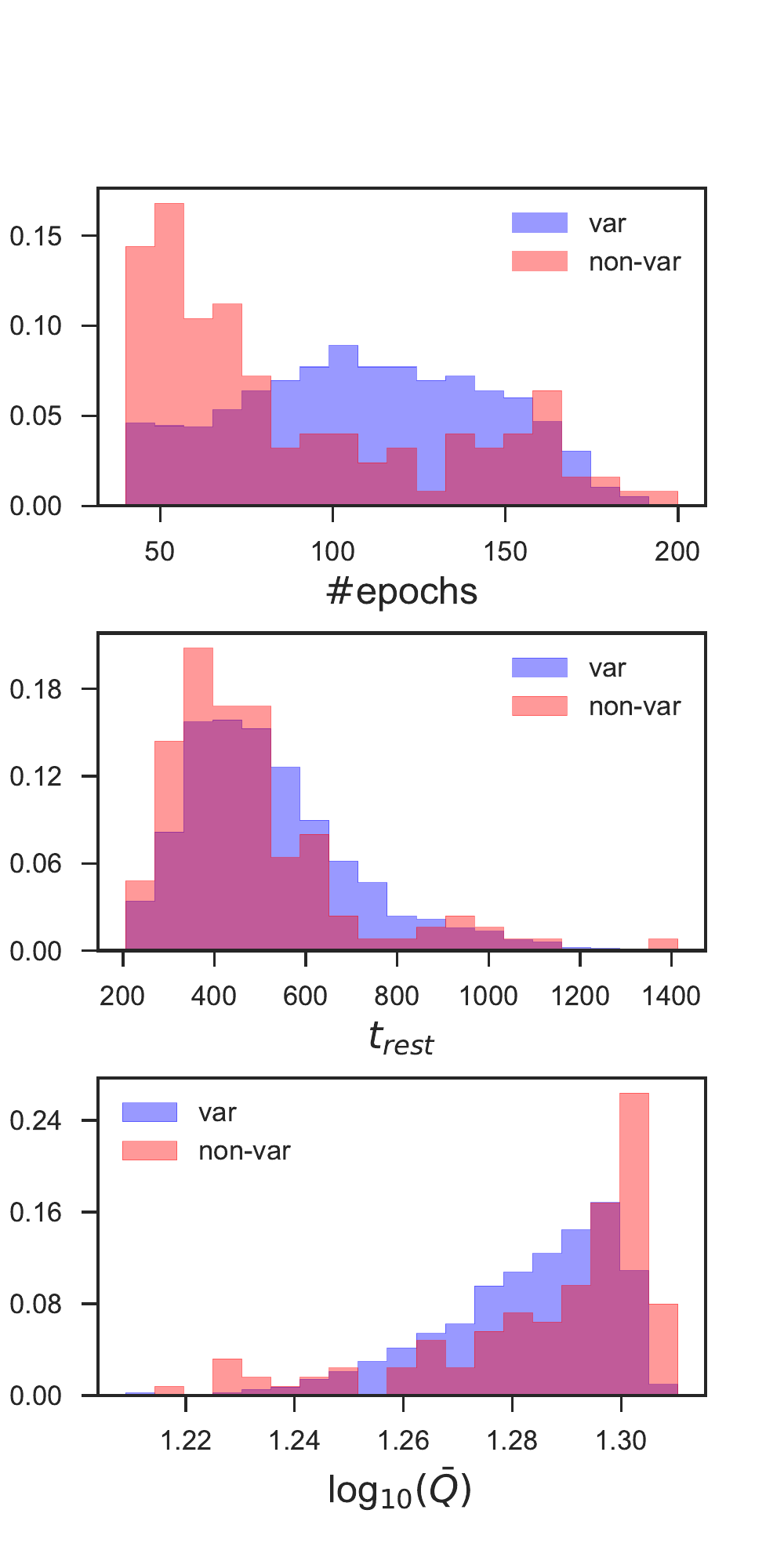}
\caption{Normalized histogram of the light curve properties (number of epochs, $t_{rest}$, and mean magnitude) of variable (blue) and non-variable (red) sources from the QUEST-SDSS sample. \label{figure:hist_var_novar_lc}}
\end{center}
\end{figure}

From the QUEST-SDSS sample, 1348 sources are variable and 125 are non-variable. Figure \ref{figure:hist_var_novar_spec} shows the normalized distribution of the different physical properties considered in this work, for the variable and non-variable sources. In the figure we can see that the distributions of L$_{5100}$, $M{_\text{BH}}$ and $L/L_{\text{Edd}}$ are similar, but for the case of non-variable sources, the distribution of $z_{spec}$ is in general shifted towards higher values of redshift, with the exception of a few sources located at low $z_{spec}$. This difference in redshift can be related with the fact that high redshift sources have shorter rest frame light curves, because of the time dilation. In Figure \ref{figure:hist_var_novar_lc} we show the normalized distribution of the light curve properties of  variable and non-variable sources from the QUEST sample. In the figure we can see that non-variable sources tend to have lower number of epochs, shorter light curves in rest frame, and fainter mean magnitudes. Therefore, in our sample, the light curve properties are more relevant for the classification of variable and non-variable sources than the physical properties of the SMBH. 

\section{Variability behavior of different classes of AGN}\label{var_classes}

\subsection{BAL QSO}\label{bal_qso}

We used the catalogs of \cite{Shen11}, \cite{Paris17a} and \cite{Paris17b} to classify 133 sources of our sample as BAL QSO. 99 of these sources have light curves with good sampling, and 86 are variable (86,9\%). In Figure \ref{figure:SF_feat} we show in red the distribution of $A$ and  $\gamma$ for the well sampled and variable BAL QSO, we can see that there is no evident difference in the SF parameters. 

In order to have a more quantitative comparison of the distribution of the SF parameters of BAL QSO and the rest of the sample, we performed a two-sample Anderson-Darling test \citep{Anderson-Darling} for the $A$ and $\gamma$ parameters. Since the Anderson-Darling test does not take into account the errors of the parameters, we only considered in the test those variable sources with a measured parameter having a signal to noise ratio higher than 3. According to the test, the distributions of the $A$ parameter are the same at a 99.5\% significance level, with a $p_{val}$ of 0.9, and the distributions of the $\gamma$ parameter are different, with a $p_{val}$ of 0.02. The difference in the distributions of $\gamma$ are related with the fact that for the rest of the sample, most of the sources are concentrated around $\gamma=0.5$, but for the case of BAL QSO, the sources are more homogeneously distributed in the SF parameter space. Nevertheless, we do not see that the values of $\gamma$ for BAL QSO are systematically different than the rest of the sample.  

\subsection{Radio Classification}\label{bal_qso}
We divided our sources as radio-loud (RL) or radio-quiet (RQ) according to their radio and optical emissions. We used data from the Faint Images of the Radio Sky at Twenty cm survey (FIRST, \citealt{Becker94}) to obtain fluxes at 20 cm of our sources. FIRST used the Very Large Array (VLA) to produce a map of the 20 cm (1.4 GHz) sky with a beam size of $5".4$ and an rms sensitivity of about 0.15 mJy beam$^{-1}$. The last version of the FIRST survey catalog (14Dec17 Version\footnote{http://sundog.stsci.edu/first/catalogs/readme.html\#coverage}), provides all the sources detected with a threshold of 1 mJy. We cross-matched our QUEST-SDSS sample with the FIRST catalog, using a radius of 1".

We classified our sources as RL and RQ using the ratio:

\begin{equation}\label{RL}
R=\frac{F_{\nu}(5 \, \text{GHz})}{F_{\nu}(4400 \, \text{\AA})}
\end{equation}

where $F_{\nu}(5 \, \text{GHz})$ is the radio flux density of the source measured at  $5 \, \text{GHz}$ and $F_{\nu}(4400 \, \text{\AA})$ is the flux density at 4400 $\text{\AA}$ \citep{Kellermann89}. We applied K-corrections to the photometry provided by the SDSS and FIRST catalogs, considering that the radio and optical emissions follow a power-law like $F \propto \nu^{-0.8}$ and $F \propto \nu^{-0.44}$ respectively. Then, we estimate $F_{\nu}(5 \, \text{GHz})$ from the measurements at 1.4 GHz provided by FIRST, and $F_{\nu}(4400 \, \text{\AA})$  from the $g$ SDSS band (4770 \AA) measurements. Therefore, the final flux values used to determine $R$ were:

\begin{eqnarray}\label{Kcorr}
F_{\nu}(5 \, \text{GHz})_{ \text{rest}}=F_{\nu}(1.4 \, \text{GHz})_{ \text{obs}} \left( \frac{1.4}{5} \right)^{0.8} (1+z)^{-0.2} \quad & \nonumber \\
F_{\nu}(4400 \, \text{\AA})_{ \text{rest}}= F_{\nu}(4770 \, \text{\AA})_{ \text{obs}} \left( \frac{6.29}{6.81} \right)^{0.44}  (1+z)^{-0.56}  \quad &
\end{eqnarray}

All the sources of the QUEST-SDSS sample are located in regions mapped by FIRST, however not all of them have a radio detection associated. 55 objects from the QUEST-SDSS sample have a FIRST counterpart. For those sources without a detection reported, we assumed that the measured flux corresponds to the detection threshold of 1 mJy. Then, we classify sources as RL if they have $R\geq10$ and are detected by FIRST, and we classify sources as RQ if they have $R<10$.

\begin{figure}
\begin{center}
\includegraphics[scale=0.4]{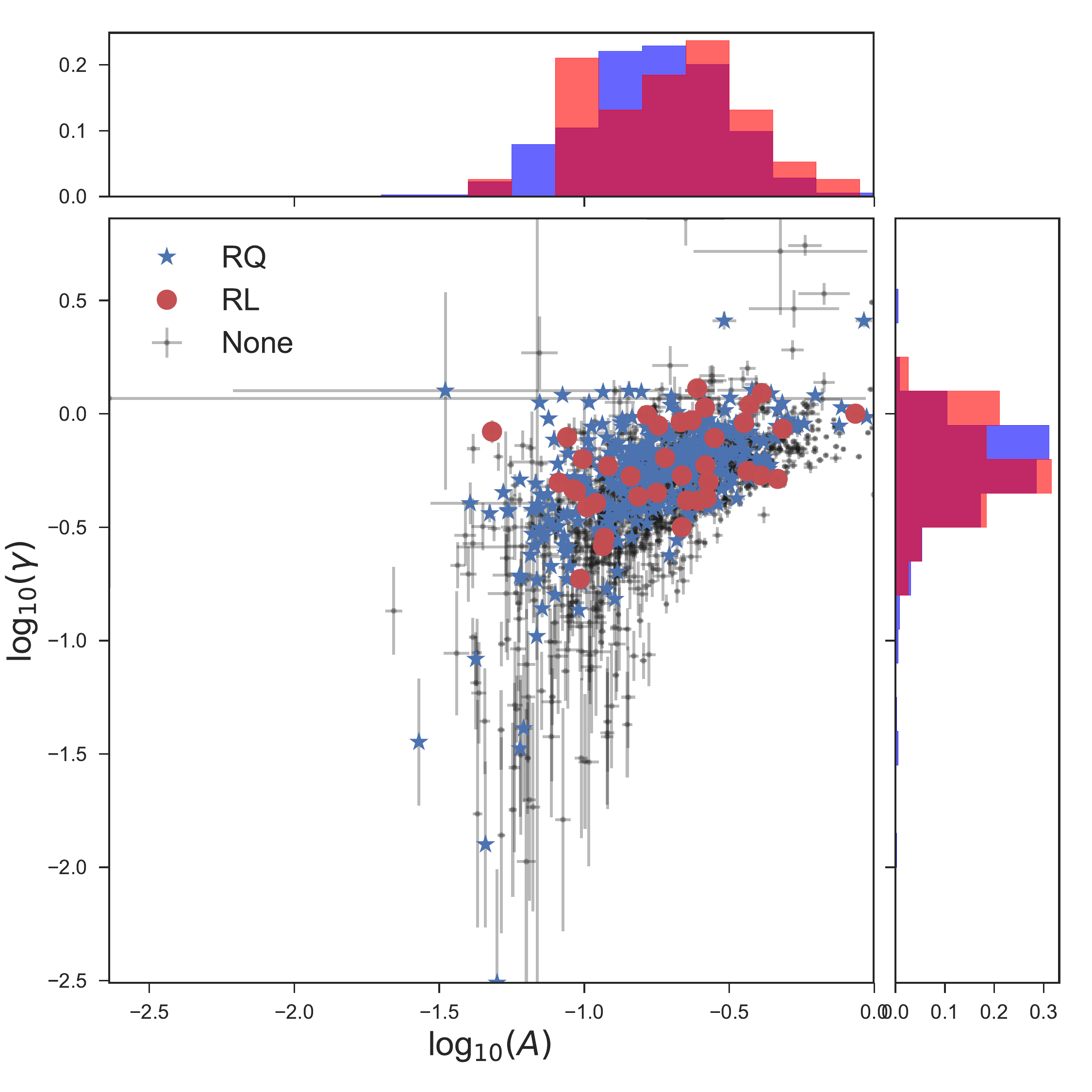}
\caption{Distribution of the SF paramenters $A$ and $\gamma$, for sources classified as RL (red circles), RQ (blue stars), and sources without radio classification (None, black dots). Along the axes we show the histograms of every parameter, for the case of RL (blue) and RQ sources (red). \label{figure:SF_RL}}
\end{center}
\end{figure}

From the QUEST-SDSS sample, 373 sources are classified as RQ and 354 are variable (94,9 \%). 48 are classified as RL and 38 are variable (79.2\%). Figure \ref{figure:SF_RL} shows the distribution of the SF parameters for RL and RQ sources, and also for sources without radio classification. In the Figure we can see that there is no evident difference in the distributions of RL and RQ sources. For a more quantitative comparison, we performed an Anderson-Darling test comparing the SF parameters distributions of the RL and RQ sources. As before, we only considered those variable sources with a measured parameter having a signal to noise ratio higher than 3. According to the test, the distributions of $A$ and $\gamma$ are the same for RQ and RL sources, with $p_{val}$ of 0.64 for $A$ and 0.22 for $\gamma$. This could imply that the radio loudness may not be relevant for the optical variability of type I AGN.

\section{Discussion and Conclusions}\label{discussion}

In section \ref{var_quest} we showed that there is a correlation between $A$ and $\gamma$, however in section \ref{sim_SF} we demonstrated that such a correlation is produced by the stochastic nature of the light curves together with the light curve sampling. We might need longer light curves to reduce this degeneracy, however, having access to long (decades of coverage) and well sampled light curves for large samples of sources is not possible currently. In section \ref{var_vs_spec} we demonstrated that $A$ anti-correlates with both $\lambda_{rest}$ and $L/L_{\text{Edd}}$, but $\gamma$ does not correlate with any of the physical parameters studied. This confirms our assumption that the correlation between $A$ and $\gamma$ is produced by the light curve properties, and by the stochastic nature of the variability. Nonetheless, the structure function is the best option that we have today to analyse typical ground-based light curves (i.e. with a few years of coverage, a few epochs, and with gaps), since other techniques, like Fourier analyses, requires well sampled light curves (i.e. with several epochs, and without gaps).

In sections \ref{spec_prop} and \ref{pca} we reported that our data set presents correlations between the SMBH physical properties. Some of these correlations are produced by selection effects, since our sample is flux limited. For example, the correlations of redshift with luminosity, BH mass, and accretion rate, are produced by the fact that at higher redshifts our sample will naturally contain sources with higher luminosities, higher SMBH masses and higher accretion rates.

The results shown in section \ref{var_vs_spec} tell us that the observed amplitude of the variability depends on two variables, $z_{spec}$ and $L/L_{\text{Edd}}$, which means that, at a fixed $z_{spec}$, sources with similar $L/L_{\text{Edd}}$ will have similar variability amplitudes. The positive correlation with redshift can be interpreted as an anti-correlation with the wavelength of rest frame emission. \cite{MacLeod10} analysed \textit{ugriz} light curves of $\sim 9000$ spectroscopically confirmed SDSS S82 quasars. Since they had multiple bands for each quasar, they could separate the dependency of the amplitude of the variability with redshift and $\lambda_{rest}$, finding an anti-correlation with $\lambda_{rest}$ and no correlation with $z_{spec}$. S17 analysed the near infrared variability of X-ray selected AGN. They also found a correlation between the amplitude of the variability and redshift. By comparing the trends between $A$ and $z_{spec}$ for two different bands (Y and J), they showed that the correlation with $z_{spec}$ is explained by an anti-correlation with the wavelength of emission. From this, and other previous results (e.g. \citealt{Kozlowski10}), we conclude that the positive correlation of $A$ with $z_{spec}$ is produced by a dependency on $\lambda_{rest}$, and is not given by evolution over cosmic time. This anti-correlation between $\lambda_{rest}$ and $A$ can be explained considering that the innermost regions of the disk can be the most variable, either intrinsically or by reprocessing. Since at shorter wavelengths a larger fraction of the disc emission is produced by the innermost region, it follows that shorter wavelengths display larger amplitudes of variability \citep{Arevalo08,Lira11,Lira15,Edelson15,Fausnaugh16}. 

We found an anti-correlation between $A$ and $L/L_{\text{Edd}}$. When we used the standard method to determine black hole masses from single epoch spectra we found a slope of $\beta_{L/L_{\text{Edd}}} = -0.19 \pm 0.02$ for the regression model with $A$ as the dependent variable, and $L/L_{\text{Edd}}$ and $z_{spec}$ as the independent variables, for the not -- C IV sample. When we apply the corrections proposed by \cite{MejiaRestrepo18a}, which intend to account for the effect of the unknown distribution of the gas clouds in the BLR, we found a slope of $\beta_{(L/L_{\text{Edd}})^C} =-0.28 \pm 0.06$.  An anti-correlation between $L/L_{\text{Edd}}$ and the amplitude of the variability has also been reported by previous works (e.g. \citealt{Wilhite08,MacLeod10,Simm16,Rakshit17}). \cite{MacLeod10} reported a power-law slope of $-0.23 \pm 0.03$. This value was calculated by binning the parameter space of $M_{\text{BH}}$ and $M_i$ (absolute magnitude), and using ensemble light curves, which can explain the small difference with the value found by us.

\cite{MacLeod10} also proposed that an additional dependency with luminosity or black hole mass is needed in order to explain their findings. Here we conclude that such a dependency is not necessary when intrinsic scatter is included in the model. In fact, in Table \ref{tab:reg_A_noCIV} we can see that the value of the intrinsic scatter found with our method is pretty stable, and that including $\text{L}_{5100}$ or $M_{\text{BH}}$ in the regression model with $z_{spec}$ and $L/L_{\text{Edd}}$ produces statistically insignificant slopes for $\text{L}_{5100}$ or $M_{\text{BH}}$. From the results of sections \ref{sim_SF} and \ref{lin_reg_A} we can say that the main contributors to this scatter are the $A-\gamma$ degeneracy (produced by the stochastic nature of the AGN variability), the definition of the SF by itself, the light-curve sampling, and the fact that light curves with coverage of a few years are not a good representation of the whole variability behavior. We could notice in section \ref{lin_reg_A} that when we performed the linear regression model with $z_{spec}$ and $L/_{Edd}$ as independent sources, selecting only those sources from the not--C IV sample whose measured values of $\gamma$ were in the range between 0.43 and 0.63, the measured scatter in the regression was reduced considerably. This confirm our assumption that the $A-\gamma$ degeneracy is one of the main contributors to the measured scatter in the regression models.

Possible interpretations of the inverse dependency of the amplitude of the variability with $L/L_\text{Edd}$ are discussed by \cite{Wilhite08}, \cite{MacLeod10}, \cite{Simm16}, and \cite{Rakshit17}. One explanation can be that $L/L_\text{Edd}$ is a proxy of the age of the AGN (e.g. \citealt{Martini03,Haas04,Hopkins05}). Sources with lower $L/L_\text{Edd}$ can suffer from a dwindling of the fuel supply, as they become old, thus, the accretion flow can be more variable, producing larger amplitude in the variability. But, the time-scales of the amplitudes measured in this work are $\sim 1$ year, and therefore, it is unlikely that the variability amplitudes observed are given by variations in the external fuel supply, which requires much longer time-scales to be effective ($10^{5}$ to $10^7$ days). 

Other possible interpretation is that sources with higher $L/L_{\text{Edd}}$ have hotter accretion disks, as predicted by classical accretion physics \citep{Shakura73}. For typical values of black hole mass and accretion rate, it is expected that the innermost part of the disk emits in the far UV. Because of its smaller size, this region is also the one showing the largest variability amplitude. For lower accretion rates however, the disk becomes cooler, and the innermost, most variable region will shift its emission from the UV to optical wavebands ($r_{\lambda} \propto M_{\text{BH}}^{2/3} (L/L_{\text{Edd}})^{1/3}\lambda^{4/3}$). This would be true regardless of whether the variation of the disk emission is produced by intrinsic processes or by reprocessing of highly variable X-ray emission by the disk surface. \cite{MacLeod10} discarded this assumption because the time-scales ($\tau$) that they measured were not in agreement with this scenario. However, they used DRW modelling to find $\tau$, while it is now clear that DRW models cannot be used to properly describe the time-scales of typical ground-based light curves (see section \ref{var_features}).  

A third possible explanation for the anti-correlation between $A$ and $L/L_{\text{Edd}}$ can be related with the positive correlation between $L/L_{\text{Edd}}$ and the ratio of the UV/optical-to-X-ray flux $(\alpha_{\text{ox}})$ reported by several studies (e.g. \citealt{Shemmer08,Grupe10,Lusso10,Jin12}). If the UV/optical variability is produced by reflection of the variable X-ray emission, then disks located in systems with higher $\alpha_{\text{ox}}$ values will receive fractionally less X-ray radiation, and therefore the amplitude of the variability detected in the UV/optical range will be small. On the other hand, for sources with lower $\alpha_{\text{ox}}$, the disk will be irradiated with more X-ray light, and therefore we will detect higher UV/optical variability amplitudes. \cite{Kubota18} developed a new spectral model for the SED of AGN that includes a hot corona, an inner warm optically thick Comptonising region and an outer disk. Considering this model, they studied the UV/optical variability resulting from the reprocessing of the rapidly variable X-ray flux. Their model predicts an anti-correlation between the amplitude of the variability and $L/L_{\text{Edd}}$. However their model also predicts  a much lower amount of UV/optical variability than what is observed by our analysis and previous studies (e.g. \citealt{MacLeod10}) at time-scales of 1 year or longer. This means that the model needs an extra source of UV/optical variability in order to explain the amplitudes observed at long time-scales, as has been found by previous analyses \citep{Krolik91,Arevalo08,Lira15,Edelson15}. Therefore, the anti-correlation between $A$ and $L/L_{\text{Edd}}$, cannot be solely explain by the correlation between $L/L_{\text{Edd}}$ and $(\alpha_{\text{ox}})$.

In this work, we also found that the logarithmic gradient of the variability ($\gamma$) does not correlate significantly with any of the physical parameter studied, and that the general distribution of $\gamma$ measured for our sample differs from the distribution of $\gamma$ obtained for light curves simulated from a DRW process. We showed in sections \ref{var_quest} and \ref{lin_reg_g} that 20,6\% of the light curves have values of $\gamma$ higher than 0.75, for which a DRW model is not appropriate to explain the variability. \cite{Kasliwal15} and \cite{Smith18} used \textit{Kepler} light curves to study whether DRW modelling is sufficient to explain the variability of light curves with high cadence. They concluded that most of the \textit{Kepler} AGN light curves analysed cannot be described by a simple DRW model. \cite{Smith18} also proposed that it is possible that DRW modelling can be correct for ground-based quasar light curves, which in general study different time regimes than \textit{Kepler}. We need larger samples of high cadence light curves, to see whether the results of \cite{Kasliwal15} and \cite{Smith18} are representative for the whole AGN population.

\acknowledgments

PS was supported by CONICYT through “Beca Doctorado Nacional, A\~no 2013” grant \#21130441, by Universidad de Chile through ``Ayudas para estad\'ias cortas de investigaci\'on para estudiantes de postgrado, convocatoria 2016-2017", and by L'Oreal Chile--UNESCO through ``For Women in Science Award, 2016". PS received partial support from Center of Excellence in Astrophysics and Associated Technologies (PFB 06). PL acknowledges Fondecyt Grant \#1161184. LCH was supported by the National Key R\&D Program of China (2016YFA0400702) and the National Science Foundation of China (11473002, 11721303). MK was supported by the Basic Science Research Program through the National Research Foundation of Korea (NRF) funded by the Ministry of Science, ICT \& Future Planning (No. NRF-2017R1C1B2002879). This work was partially funded by the CONICYT PIA ACT172033.

Funding for the SDSS and SDSS-II has been provided by the Alfred P. Sloan Foundation, the Participating Institutions, the National Science Foundation, the U.S. Department of Energy, the National Aeronautics and Space Administration, the Japanese Monbukagakusho, the Max Planck Society, and the Higher Education Funding Council for England. The SDSS Web Site is http://www.sdss.org/.

The SDSS is managed by the Astrophysical Research Consortium for the Participating Institutions. The Participating Institutions are the American Museum of Natural History, Astrophysical Institute Potsdam, University of Basel, University of Cambridge, Case Western Reserve University, University of Chicago, Drexel University, Fermilab, the Institute for Advanced Study, the Japan Participation Group, Johns Hopkins University, the Joint Institute for Nuclear Astrophysics, the Kavli Institute for Particle Astrophysics and Cosmology, the Korean Scientist Group, the Chinese Academy of Sciences (LAMOST), Los Alamos National Laboratory, the Max-Planck-Institute for Astronomy (MPIA), the Max-Planck-Institute for Astrophysics (MPA), New Mexico State University, Ohio State University, University of Pittsburgh, University of Portsmouth, Princeton University, the United States Naval Observatory, and the University of Washington.

\bibliography{bibliography}

\end{document}